\documentclass[aps,prx,twocolumn,superscriptaddress]{revtex4-2}
\usepackage{amssymb}
\usepackage{suffix}
\usepackage{float}
\usepackage{mathtools}
\usepackage[utf8]{inputenc}
\usepackage{booktabs}
\usepackage{cases}
\usepackage[multiple]{footmisc}
\usepackage{dcolumn}
\usepackage{color,soul}
\usepackage{rotating}
\usepackage{perpage}
\usepackage{xcolor}
\usepackage{soul}
\usepackage{tikz}
\usepackage[T1]{fontenc}
\usepackage{etoolbox}
\usepackage{graphics}
\usepackage{siunitx}
\usepackage{hyperref}
\usepackage{float}	
\usepackage{collref}
\usepackage{multirow}
\usepackage{mathtools}
\usepackage{bm}
\usepackage{url}

\usepackage{tikz}
\usepackage{tikz-3dplot}
\usepackage{graphicx}
\usepackage{epsfig}
\usepackage{subfigure}
\usepackage{epstopdf}
\usepackage{nccmath}
\newcommand{\MY}[1]{\textcolor{red}{\fbox{MY} {\sl#1}}}
\newcommand{\NM}[1]{\textcolor{violet}{\fbox{NM} {\sl#1}}}
\newcommand{\MK}[1]{\textcolor{blue}{\fbox{MK} {\sl#1}}}

\AtBeginDocument{
	\setlength\abovedisplayskip{4pt}
	\setlength\belowdisplayskip{4pt}}

\newcounter{myparagraphs}

\begin{document} 
	\title{Spin high-harmonic generation through terahertz laser-driven phonons}
	\author{Negin Moharrami Allafi}
	\email{nmoharramiallafi@gradcenter.cuny.edu}
	\affiliation{Physics program and Initiative for the Theoretical Sciences, The Graduate Center, CUNY, New York, NY 10016, USA}
	
	\author{Michael H. Kolodrubetz}
	\email{mkolodru@utdallas.edu}
	\affiliation{Department of Physics, The University of Texas at Dallas, Richardson, Texas 75080, USA}

 \author{Marin Bukov}
	\email{mgbukov@pks.mpg.de}
	\affiliation{Max Planck Institute for the Physics of Complex Systems, N\"othnitzer Str.~38, 01187 Dresden, Germany}

 \author{Vadim Oganesyan}
	\email{vadim.oganesyan@csi.cuny.edu}
	\affiliation{Physics program and Initiative for the Theoretical Sciences, The Graduate Center, CUNY, New York, NY 10016, USA}
	\affiliation{Center for Computational Quantum Physics, Flatiron Institute, 162 5th Avenue, New York, NY 10010, USA}    
	\affiliation{Department of Physics and Astronomy, College of Staten Island, CUNY, Staten Island, NY 10314, USA}
 
	\author{Mohsen Yarmohammadi}
	\email{mohsen.yarmohammadi@utdallas.edu}
	\affiliation{Department of Physics, The University of Texas at Dallas, Richardson, Texas 75080, USA}
	
	\date{\today}
	\begin{abstract}
		In the realm of open quantum systems, steady states and high-harmonic generation (HHG) existing far from equilibrium have become core pillars of ultrafast science. Most solid-state research explores charge HHG with limited investigations into spin degrees of freedom. In this study, we theoretically address spin HHG in the steady state resulting from the terahertz laser-driven spin-phonon coupling in a dissipative dimerized spin-1/2 chain. Instead of directly driving spins using time-dependent magnetic fields, we employ the magnetophononic mechanism, where the laser first drives the lattice, and then the excited lattice subsequently drives the spins. We investigate the role of various model parameters for optimizing HHG. Increasing the laser's amplitude amplifies spin HHG beyond the perturbative regime, enhancing both harmonic amplitudes and orders. We find that configuring the drive frequency far below the spin band yields the highest harmonic order. Additionally, we provide a theory matching the numerical results under weak spin-phonon coupling and propose an experimental procedure to probe the emission spectrum of spin HHG.
	\end{abstract}
	
	\maketitle
	
	{\allowdisplaybreaks
		\section{Introduction}\label{s1}
		Driving an open quantum system involves non-equilibrium dynamics where the system interacts with its environment to reach a steady state. Steady states play vital roles in quantum information encoding~\cite{PhysRevA.95.042302}, simulations/algorithms~\cite{PhysRevA.93.032306,PhysRevLett.130.240601}, Floquet engineering~\cite{doi:10.1126/sciadv.abb4019,PhysRevResearch.2.043289}, and studying phase transitions~\cite{PhysRevLett.105.015702,Chertkov2023,PhysRevB.108.L140305,yarmohammadi2024ultrafast}. With continuing advancements in lasing technology~\cite{kampfrath2013resonant,kruchinin2018colloquium}, there is also significant interest in high-harmonic generation (HHG) arising from steady states~\cite{RevModPhys.72.545,Krausz_2016,ghimire2019high,li_attosecond_2020,yu2019high,goulielmakis_high_2022}. HHG involves nonlinear interactions between a pump laser and material, emitting integer multiples of the drive frequency, which has applications in various scientific and industrial domains such as spectroscopy~\cite{marangos2016development,plotzing2016spin,lloyd20212021}, microscopy~\cite{CAMPAGNOLA2002493,eschen2022material}, lensless imaging~\cite{witte2014lensless,gardner2017subwavelength}, high-precision measurements ~\cite{RevModPhys.78.1297} as well as ultrafast dynamics~\cite{baker2006,uiberacker2007attosecond} and attosecond sciences~\cite{RevModPhys.81.163,li_attosecond_2020,shi2020attosecond}.
		
		The concept of HHG, initially focused on laser-driven atomic and molecular gases~\cite{McPherson:87,MFerray_1988,PhysRevLett.70.774,PhysRevLett.70.766}, has recently gained attention in solids due to advancements in laser technology~\cite{ghimire_observation_2011,ghimire2012generation, ndabashimiye2016solid}. Solids, with their variable band structure, high atomic density, and crystal lattice, offer insights into material characteristics~\cite{Vampa_2017,vampa2014theoretical,mcdonald2015interband}. Various solid-state materials including semiconductors~\cite{freeman2022high}, topological insulators~\cite{baykusheva2021strong,schmid2021tunable,bai2021high,ma2022role,graml2023influence}, Mott insulators~\cite{udono2022excitonic,murakami2022anomalous}, and low-dimensional materials~\cite{rana2022generation,cha2022gate,gnawali2022ultrafast} have been investigated for harmonic generation. This has provided avenues to probe ultrafast dynamics in solids and for the all-optical reconstruction of the band structure~\cite{PhysRevLett.115.193603,lanin2017mapping}.
		
		Most studies on solid-state HHG are focused on electronic systems, leveraging charge degrees of freedom~\cite{schubert2014sub, hohenleutner2015real,yu2019high,ghimire2019high}. However, high harmonics can also originate from spin systems, although research in this area is limited. Addressing spin HHG has significant potential for applications in spintronics~\cite{RevModPhys.76.323,HIROHATA2020166711,YUAN20221}, offering faster computation and higher efficiency with reduced power consumption for tasks like data processing. While recent HHG studies hint at spin system involvement~\cite{zhang2018generating,PhysRevB.102.081121,PhysRevB.99.184303,kanega2021linear,PhysRevB.100.214424}, challenges arise in tuning magnetic field strength for spin excitation, potentially complicating experimental setups. 
		
		In this paper, we propose a comprehensive theoretical framework for generating high harmonics from gapped quantum magnets. Rather than directly coupling the laser's magnetic field to the spin system, we employ the terahertz ``magnetophononics'' mechanism~\cite{PhysRevMaterials.2.064401,PhysRevB.103.045132,PhysRevB.107.174415,PhysRevB.107.184440,PhysRevB.107.184440,yarmohammadi2023laserenhanced}, where the terahertz laser's electric field interacts with phonons, first, and second drives the spin sector via the phonons, see Fig.~\ref{fig1}. This mechanism incorporates the lattice structure inherent in solid-state materials, providing a means to manipulate the spin system. 
		
		We develop a model for a dimerized spin-1/2 chain like CuGeO$_3$~\cite{PhysRevB.63.094401,PhysRevB.103.045132,PhysRevB.107.174415,Chen2021}, where the driven phonon couples to both intradimer and interdimer interactions, forming strong nonlinear inreractions through the so-called $J$-$J'$ model. This microscopic Hamiltonian governs spin and phonon dynamics alongside a phenomenological Lindbladian to model bath, which we then approximate to enable numerical simulations using a mean-field-decoupled equations of motion formalism. 
		Exploring a range of model parameters, our goal is to optimize spin HHG generation. We aim to investigate spin HHG using a physically motivated observable that measures the power emitted by the spins, which facilitates the monitoring of the HHG profile across various parameters.
		
		We observe that increasing the laser's amplitude drives spin HHG beyond the perturbative regime, leading to enhanced harmonic amplitudes and orders. Interestingly, we identify that setting the drive frequency far below the spin band results in the highest attained harmonic order. Furthermore, we develop a theoretical framework in cases of weak spin-phonon coupling that closely matches numerical simulations. To detect this phenomenon, we propose an experimental approach to investigate the emission spectrum of spin HHG. These findings, with their potential implications for spintronics, highlight the technological progress enabled by HHG. 
		
		The paper is structured as follows. In Sec.~\ref{s2}, we present the Hamiltonian model describing a dimerized spin-1/2 chain interacting with laser-driven phonons. Using the Lindblad quantum master equation, we derive the time evolution of coupled particles in the model. Section~\ref{s3} presents our numerical and analytical findings and discusses potential experimental observations. Finally, we summarize the paper in Sec.~\ref{s4}.	
		
		\section{Model and methods}\label{s2}
		
		We start with an illustration of a dimerized spin-1/2 chain, as exemplified by Cu-O-Cu exchange paths in CuGeO$_3$, subject to a continuous terahertz laser field. The setup is illustrated in Fig.~\ref{fig1}. Each copper ion accommodates two spins, and the vibrations of the oxygen ions, primarily induced by the laser field, result in the excitation of spins through spin-phonon coupling. 
		
		\subsection{Hamiltonian: spins, phonon and couplings}
		\begin{figure}[t]
			\centering
			\includegraphics[width=1\linewidth]{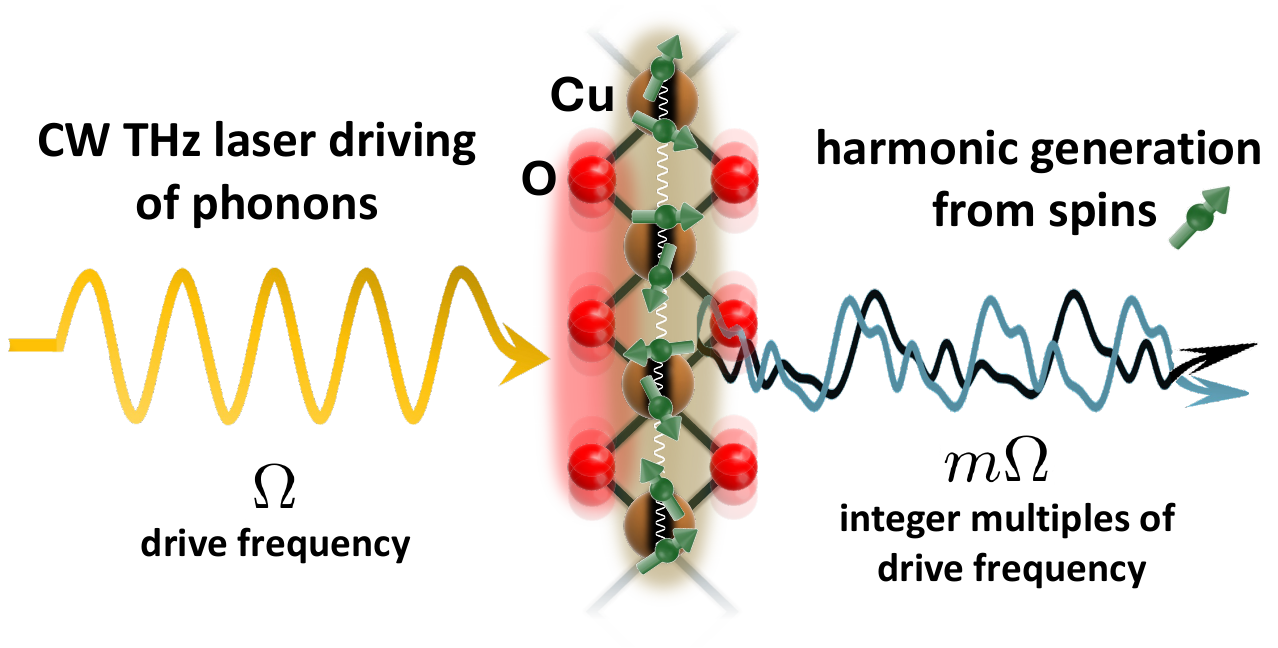}
			\caption{\textbf{Schematic of spin high-harmonic generation}. Spin HHG stimulated by terahertz laser-driven phonons, in a gapped quantum magnet CuGeO$_3$, specifically a dimerized spin-1/2 chain of Cu-O-Cu exchange path. Phonons are mainly produced by the oscillation of driven oxygen ions. Each dimer consists of two spins 1/2 of Cu$^{2+}$ ions.}
			\label{fig1}
		\end{figure}
		The total Hamiltonian constitutes of four terms, namely the dimerized spin chain $\mathcal{\bf H}_{\rm s}$, the phonon system $\mathcal{\bf H}_{\rm p}$, the laser-phonon coupling $\mathcal{\bf H}_{\rm l}$, and the spin-phonon coupling~(SPC) $\mathcal{\bf H}_{\rm sp}$~(we set $\hbar = 1$):
		\begin{subequations}
			\begin{align}
				&\mathcal{\bf H}_{\rm s} ={}\sum^L_{\ell=1} \left[ J \,\vec{\bf S}_{1,\ell} \cdot \vec{\bf S}_{2,\ell} + J' \,\vec{\bf S}_{2,\ell} \cdot \vec{\bf S}_{1,\ell+1}\right], \label{eq_01b}\\
				&\mathcal{\bf H}_{\rm p}+\mathcal{\bf H}_{\rm l}={}\omega_0 {\bf a}^
				\dagger {\bf a} + L E(t) {\bf x}\,, \\
				&\mathcal{\bf H}_{\rm sp} = {}{\bf x}\sum^L_{\ell=1} \left[  g\,\vec{\bf S}_{1,\ell} \cdot \vec{\bf S}_{2,\ell} + g'\,\vec{\bf S}_{2,\ell} \cdot \vec{\bf S}_{1,\ell+1}\right]\, ,
			\end{align}
		\end{subequations}
		where bold letters are generally reserved for quantum operators. Here ${\bf a}$  is the zero-momentum phonon mode, which is the only optical (Einstein) phonon that couples to the spins $\vec{\bf S}$; ${\bf x}=({\bf a}+{\bf a}^\dagger)/\sqrt{L}$ is its displacement. Parameters $J~( J'= 0.5 J)$ refers to the intradimer (interdimer) spin-spin coupling strength and $L$ is the number of dimers, see Fig.~\ref{fig1}. Furthermore, $\omega_0$ and $\{g,g'\}$ are the phonon frequency and the strength of SPC to the $\{J,J'\}$ bond, respectively. The laser field is assumed to be a classical monochromatic electric field $E(t)=\mathcal{A} \cos(\Omega\,t)$ with frequency $\Omega$ and uniform coupling to infrared-active phonons, i.e., only to the uniform displacement of (primarily) the oxygen sublattice. The amplitude of the drive $\mathcal{A}$ is a product of the laser's electric field strength and the dipole moment of this phonon mode~\cite{PhysRevMaterials.2.064401}.
		
		
		We include magnetoelastic effects by assuming that both intradimer and interdimer bonds are affected linearly by lattice distortions along the chain~\cite{PhysRevB.59.14356}; we name it the $J$-$J'$ model. Although the linear SPC Hamiltonian can be justified by the fact that the expected lattice distortions in CuGeO$_3$ are rather small~\cite{PhysRevB.56.R11357}, this Hamiltonian does not describe realistic phonon dynamics in detail, e.g., their nonlinearities which would redistribute the energy away from the pumped mode. This physics is captured phenomenologically using the Lindblad formalism (see below). 
		
		\subsection{Triplon decoupling}		
		We employ the bond-operator representation to rewrite the spin operators in a dimer using bosonic ``triplons'' (represented by ${\bf t}$ and ${\bf t}^\dagger$ below) 
		\cite{PhysRevB.41.9323,PhysRevB.49.8901}. 
		We set the singlet state as vacuum and focus on the triplet excitations through 
		\begin{subequations}
			\begin{align}
			{\bf S}^{\alpha}_{1,\ell} = {} & \frac{1}{2}({\bf t}_{\alpha,\ell}+{\bf t}^\dagger_{\alpha,\ell} 
		- i \sum_{\beta \zeta}\epsilon_{\alpha \beta \zeta} {\bf t}^\dagger_{\beta,\ell}\,{\bf t}_{\zeta,\ell})\, ,\\
		{\bf S}^{\alpha}_{2,\ell} = {} & \frac{1}{2}(-{\bf t}_{\alpha,\ell}-{\bf t}^\dagger_{\alpha,\ell} 
		- i \sum_{\beta \zeta}\epsilon_{\alpha \beta \zeta} {\bf t}^\dagger_{\beta,\ell}\,{\bf t}_{\zeta,\ell})\, ,
						\end{align}
		\end{subequations}where $\epsilon$ is the totally antisymmetric tensor considering different flavors $\alpha = \{x,y,z\}$. 
		Note, that staggered (antisymmetric) dimer spin operators ${\bf S}^{\alpha}_{1,\ell}-{\bf S}^{\alpha}_{2,\ell}$ are linear in triplons while uniform (symmetric) combinations are quadratic, e.g., ${\bf S}^{z}_{1,\ell}+{\bf S}^{z}_{2,\ell}=i ({\bf t}^\dagger_{x,\ell}\,{\bf t}_{y,\ell}-{\bf t}^\dagger_{y,\ell}\,{\bf t}_{x,\ell})$ and become important in the presence of a magnetic field.
		In what follows we neglect 
		all triplon-triplon interactions (justified by the low occupation of these modes in the NESS assuming a zero temperature, and the absence of a magnetic field)
		and employ the Bogoliubov transformation 
		\begin{subequations}
			\begin{align}
				&{\bf t}_{k,\alpha} = {} \tilde{{\bf t}}_{k,\alpha} \cosh(\theta_k) +\tilde{{\bf t}}^{\,\dagger}_{-k,\alpha} \sinh(\theta_k)\,, \\
				&\mathcal{\bf H}_{\rm s}\approx -\frac{3}{4}\,\big[L\,J + J' \sum_{k} \cos(k)\big] + \sum_{k,\alpha} \varepsilon_k \tilde{\bf t}^{\,\dagger}_{k,\alpha} \tilde{\bf t}_{k,\alpha}\,, \label{eq_1b}\\
				&\varepsilon_k = J \sqrt{1 - \frac{J' }{J}\cos(k)} \,,\label{eq:dispersion}
			\end{align}
		\end{subequations}
		to diagonalize the spin Hamiltonian (truncated to quadratic order in triplon field operators), where $\varepsilon_k$ represents the triplon dispersion and $\exp(-2\theta_k) = \varepsilon_k/J$.

We then proceed to rewrite and approximate the spin-phonon interactions through mean-field decoupling
		\begin{subequations}
			\begin{align}
				&\mathcal{\bf H}_{\rm sp,s} \approx {\bf x} \sum_{k,\alpha} \Big[A_k\,\tilde{{\bf t}}^{\,\dagger}_{k,\alpha} \tilde{{\bf t}}_{k,\alpha} + \frac{B_k}{2}\,(\tilde{{\bf t}}^{\,\dagger}_{k,\alpha} \tilde{{\bf t}}^{\,\dagger}_{-k,\alpha}+\text{h.c.})\Big] \, ,\\
				&
				\label{eq_6_gk}
				A_k \equiv {} g  \mathcal{R}_k - \frac{g' J}{J'} \mathcal{S}_k \quad , \quad
				B_k \equiv {} \Big(g-\frac{g' J}{J'}\Big) \mathcal{S}_k\, ,
				\\
				&\mathcal{R}_k   \equiv \frac{J}{\varepsilon_k}[1-\frac{J' }{2J} \cos(k)] \quad , \quad \mathcal{S}_k  \equiv \frac{J'}{2\varepsilon_k} \cos(k)\,.
			\end{align}
		\end{subequations}

		\subsection{Bath coupling}       
		We treat the bath -- consisting of undriven phonons in thermal equilibrium -- via a phenomenological Lindblad formalism. The Lindblad quantum master equation~\cite{breuer2007theory,lindblad1976} approximates an evolution of an arbitrary operator ${\bf O}$ in an open quantum system
		with a time-local norm preserving formalism
		\begin{align}
			\label{eq_25} 
			\dot {\bf O} (t) = i [\mathcal{\bf H}(t),{\bf O}]+ \sum_{n} \gamma_{n} \big[{\bf L}_{n}^{\dagger}{\bf O}{\bf L}_{n}  -\frac{1}{2}\{{\bf L}_{n}^{\dagger}{\bf L}_{n},{\bf O}\}\big],
		\end{align} 
		where $\mathcal{\bf H}(t)$ is the total time-dependent Hamiltonian, 
		${\mathcal{\bf L}}_{n}=\{{\bf t}_{k,\alpha},\,{\bf t}^\dagger_{k,\alpha},\,{\bf a},\,{\bf a^\dagger}\}$ are the so-called jump operators in the reduced system's Liouville space and 
  $\gamma_{n}= \{\gamma_{\rm s}[n(\varepsilon_k) +1],\,\gamma_{\rm s}n(\varepsilon_k),\, \gamma_{\rm p}[n(\omega_0)+1],\,\gamma_{\rm p}n(\omega_0)\}$ are phenomenological damping parameters. The $n(\varepsilon_k)$ and $n(\omega_0)$ terms are representing the bosonic occupation numbers of triplon and phonon modes, respectively. 
  In very low temperature approximation $T \to 0$, which is the case we assume, $n(\varepsilon_k)$ and $n(\omega_0)$ terms vanish.
  \iffalse, and the damping parameters are reduced to $\gamma_{n}= \{\gamma_{\rm s},\, 0,\, \gamma_{\rm p},\,0\}$ for ${\mathcal{\bf L}}_{n}$ jump operators}\fi 
  For simplicity, we assume 
		the same phonon damping when phonons are coupled to the $J$ and $J'$ bonds. In addition to being natural due to the phononic bath and uncaptured phonon non-linearities, dissipation becomes important for balancing the drive to enable a nontrivial NESS. 
		
		\subsection{Equations of motion}        
		Now, we turn to define the physical observables for both phonon and spin systems. The common observables in a phononic system at any time $t$ are displacement, momentum, and phonon number, with triplon occupation and pairing~(excitation) in the spin system, rounding out the set
		\begin{subequations}
			\begin{align}
				\label{eq_23}
				&q(t)=\langle {\bf x}\rangle(t)\,,\\
				&p(t)=\frac{i}{\sqrt{L}} \langle {\bf a}^\dagger - {\bf a} \rangle(t)\,, \\
				&N(t) = \frac{1}{L} \langle {\bf a}^\dagger {\bf a} \rangle(t)\,, \\
				&n_k(t) = \sum_\alpha \langle \tilde{\bf t}^{\,\dagger}_{k,\alpha} \tilde{\bf t}_{k,\alpha}  \rangle (t)\,,\\
				&z_k(t) =  \sum_\alpha \langle \tilde{\bf t}^{\,\dagger}_{k,\alpha} \tilde{\bf t}^{\,\dagger}_{-k,\alpha}\rangle(t) \, ,
				\label{eq_24b}
			\end{align}
		\end{subequations}
		which are all real except $z_k(t)$. For future convenience, we additionally introduce $z_k(t)\equiv v_k(t) + i w_k(t)$. 

	Considering all terms of the total Hamiltonian at zero temperature, treating the triplons as ordinary bosons, and ${\bf L}_n = \{{\bf a},{\bf a^\dagger}\}$ operators in Eq.~\eqref{eq_25}, one achieves 
	\begin{subequations} \label{eq_29} 
		\begin{align}  
			\dot q(t) = {} &\omega_{0}p(t) -  \frac{\gamma_{\rm p}}{2} q(t)\, ,\label{eq_29a}\\ 
			\dot p(t) = {} &- \omega_{0} q(t) -2 \widetilde{E}(t) -  \frac{\gamma_{\rm p}}{2} p(t) \, ,\label{eq_29b}\\
			\dot N(t) ={} &  -\widetilde{E}(t) p(t) - \gamma_{\rm p} N(t) \, ,\label{eq_29c}
		\end{align} 
	\end{subequations}
	where the effective laser field~(after including feedback from the couplings $g$ and $g'$) acting on the phonon is given by\begin{equation}	\label{eq_30_new}
		\widetilde{E}(t) = {} E(t) + \frac{1}{L} \sum_{k} A_k n_k(t) + \frac{1}{L} \sum_{k} B_k v_k(t) \, .
	\end{equation}At this point, it is worth noting that high harmonics are generated due to such dressing effects. The weak damping of spin excitations due to a phononic bath allows us to treat the magnetic excitations as weakly damped oscillators. Therefore, for the spin or triplon system, applying ${\bf L}_{n}= \{{\bf t}_{k,\alpha},{\bf t}^\dagger_{k,\alpha}\}$, we obtain
	\begin{subequations} \label{eq_36} 
		\begin{align} 
			&\dot n_k(t) =  {} 2\,B_k \,q(t)\,w_k(t)- \gamma_{\rm s}n_k(t) \,,\label{eq_36a}\\
			&\dot v_k(t) = {}  - 2  \left[\varepsilon_k + A_k \,q(t)\right]\, w_k(t) -\gamma_{\rm s}\, v_k(t) \,,\label{eq_36b}\\
			&\dot w_k(t) = {}  2 \left[\varepsilon_k + A_k \,q(t)\right]\,v_k(t) + 2 \,B_k\, q(t)\, [n_k(t) + {\textstyle \frac{3}{2}}] \notag \\ {} &\hspace{1.2cm} - \gamma_{\rm s}\, w_k(t)\, .\label{eq_36c}
		\end{align} 
	\end{subequations}

	\section{Results and discussion \label{s3}}
We numerically solve the coupled differential Eqs.~\eqref{eq_29} and~\eqref{eq_36} with the number of dimers set to $L=701$ -- we found that was sufficiently large for convergence. Furthermore, we have set the dimerization factor to $J'/J = 0.5$, while we have also investigated smaller and larger factors in Appendix~\ref{apa} to validate that a narrower spin band tends to favor more harmonics. For Fourier analysis, we trimmed the data to focus on the NESS and performed a fast Fourier transform on a sufficiently large window of temporal data. 
	
	The effective classical model derived in the previous section is still too complex to immediately identify its interesting dynamical regimes. Its nine-dimensional parameter space is spanned by laser amplitude and frequency ($\mathcal{A}, \Omega$), phonon frequency and damping ($\omega_0, \gamma_{\rm p}$), spin-spin, spin-phonon interactions and spin damping ($J, J', g, g', \gamma_{\rm s}$). We narrow and organize our exploration guided by physical and practical considerations: \begin{itemize}
		\item The laser amplitude $\mathcal{A}/\gamma_{\rm p}<1$ is considered to be a reasonable choice for our spin model with dimerization at low-temperatures~\cite{PhysRevB.51.6777,PhysRevLett.70.3651} to avoid system melting and decoherence effects~\cite{osti_6089125,PhysRevB.103.045132}, see also Appendix \ref{sec:offresonantetc};
		\item The terahertz laser-driven phonon is weakly coupled to the phononic bath, thus, we may set the $\gamma_{\rm p} < 0.1\, \omega_0$ to be a few percent of the phonon energy for most cases;
		\item Damping of the spin/triplon excitations~($\gamma_{\rm s}$) is generally weaker than phonon damping, thus, we set $\gamma_{\rm s} < \gamma_{\rm p}$~(see Appendix~\ref{apb} for additional exploration of $\gamma_{\rm p}$ and $\gamma_{\rm s}$ dependence);
		\item We choose to set $g/J  > g'/J'$ to avoid vanishing $B_k$ in Eq.~\eqref{eq_36a} responsible for spin HHG through SPC.
	\end{itemize}
	
	In what follows, we primarily assume that the pump field resonantly drives the chain at $\Omega = \omega_0$ to ensure that the phonon is strongly excited. We also discuss the off-resonance regime $\Omega \neq \omega_0$ in Appendix~\ref{sec:offresonantetc}.
	
	\subsection{Spin emitted power}
	Under the conditions of terahertz continuous driving, we expect the NESS to form on picosecond timescales, largely set by $\gamma_n$, the coupling to the environment. 
	The structure of the NESS itself is determined by the interplay of spin and spin-phonon interactions and may be complex.  In what follows we seek to identify simple quantities to interpret and possibly measure in the experiment. One such quantity is the displacement $q(t)$ of the phonon mode which should develop its high-frequency profile 
	through interactions with spins and be detected straightforwardly via its dipole radiation. However, the HHG profile originating from the phonon sector appears weak, as a significant portion of the input energy is directly dissipated into the bath~\cite{PhysRevB.103.045132}.
 
 Conversely, the remaining energy is transferred to the spin-phonon coupling, thereby amplifying the spin HHG signal. Hence, we focus on the power transmitted by the spins into the environment. At the level of approximations above, spin energy at any time $t$ is simply $\sum_k \varepsilon_k n_k(t)$ and therefore, Eq. \eqref{eq_36a} is simply a statement of power balance in the steady state (on the right-hand side), i.e., power absorbed from the phonon (first term) cancels exactly power radiated into the bath (second term). Considering $M = \rho d$, the mole amount of our sample per cm$^2$ with the molar density of $\rho \approx 0.03$ mol.cm$^{-3}$~\cite{Ecolivet_1999} and thickness $d \approx 10$ nm, the emitted spin power per cm$^2$ reads
	
	\begin{equation}\label{eq:power}
		\mathcal{P}_{{\rm t \to b}}(t) = \frac{M n_{\rm a}\gamma_{\rm s}}{2 L} \sum_k \varepsilon_k n_k(t)\, ,
	\end{equation}where $n_{\rm a}$ is the Avogadro's number. 
	
	The contribution of higher frequency components to the energy flow can be assessed by the following Fourier series in a NESS driven by any frequency $\Omega$:
	\begin{equation}\label{eq_3}
		\mathcal{P}_{{\rm t \to b}}(t) = \sum_m \mathcal{P}_{{\rm t \to b},m} \,e^{i m \,\Omega \,t}\, ,
	\end{equation}where integer $m$ is the harmonic number; $m = 0$ refers to the time-average of signals in the NESS. Any of the $m\neq 0$ components has both real and imaginary parts; we will primarily be interested in their magnitude and ignore the phase. We consider powers above an arbitrary threshold, approximately 10$^{-8}$ W/cm$^2$. Importantly, this power profile can also be detected electromagnetically with the introduction of uniform magnetic (Zeeman) field as the oscillations of triplon interaction energy couple into oscillations of Zeeman energy and with its magnetization along the field direction~\cite{PhysRevLett.123.197204,Tashiro2015,Chen2021}.
	
	\subsection{Spin HHG signal in the NESS}
	\begin{figure}[t]
		\centering
		\includegraphics[width=0.9\linewidth]{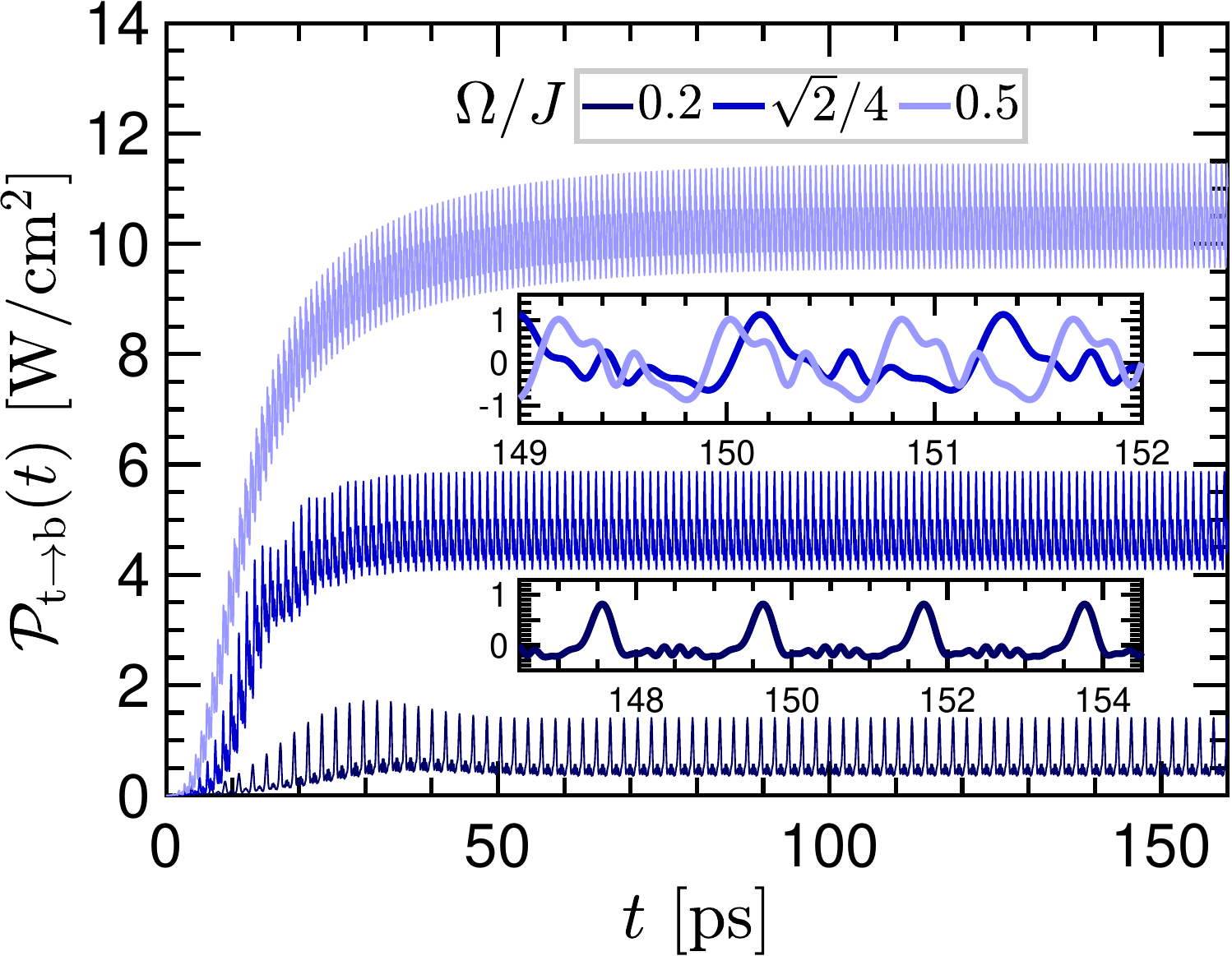}
		\caption{\textbf{Spin HHG signal in time}. Time evolution of the energy outflow from the spin system into the phononic bath, for the $J\mbox{-}J'$ model, starting from $t=0$ until it reaches NESS at long enough time, represented for $\Omega/J=0.2$, $\sqrt{2}/4$, and $0.5$. The inset panels show the complex signal patterns formed in the NESS and for comparison, average signal values are deducted. For all three frequencies, we set $\Omega=\omega_0$, $\gamma_{\rm p}/\omega_0=0.06$, $\gamma_{\rm s}/J=0.011$, $\mathcal{A}/ \gamma_{\rm p}=0.5$, and spin-phonon coupling strengths are $g/J=0.5$ and $g'/J'=0.3$.} 
		\label{fig2}
	\end{figure}
	To address harmonic superpositions, we first establish the NESS in the time domain once sufficient time has passed. Figure~\ref{fig2} illustrates the time evolution of $\mathcal{P}_{{\rm t \to b}}(t)$, representing the energy flow from triplons into the bath as defined in Eq.~\eqref{eq:power}, for three different low frequencies, $\Omega/J=0.2$, $\sqrt{2}/4$, and 0.5. The average values of the NESS are subtracted for easier comparison in the inset panels. As shown, the electric field pumps $\mathcal{P}_{{\rm t \to b}}(t)$ at $t=0$ into the NESS, where periodic oscillations occur around a constant value, equal to the absolute value of the $0$-th order harmonic, $|\mathcal{P}_{{\rm t \to b},0}|$. The timescale for the spin system to reach NESS for the presented frequencies is approximately $t\approx 100$ ps. The emergence of higher harmonic orders in the spin HHG spectra is evident due to complex patterns in the NESS, which is heavily influenced by incorporating the coupling of driven phonon to both the $J$ and $J'$ bonds, compared to previous works~\cite{PhysRevB.103.045132,PhysRevB.107.174415}. 
	To be more specific, the $A_k$ and $B_k$ terms given by Eq.~\eqref{eq_6_gk} allow the integration of the combined effects of both SPC terms, as a bundle, into the equations of motion. 
	These bundle terms blend into the spin sector, through the effective drive term $\widetilde{E}(t)$ in Eq.~\eqref{eq_30_new}, enabling the indirect stimulation of the spins along with the laser drive field.

	\subsection{Spin-phonon coupling effects on spin HHG and weak coupling theory}

\begin{figure}[t]
		\centering
		\includegraphics[width=0.85\linewidth]{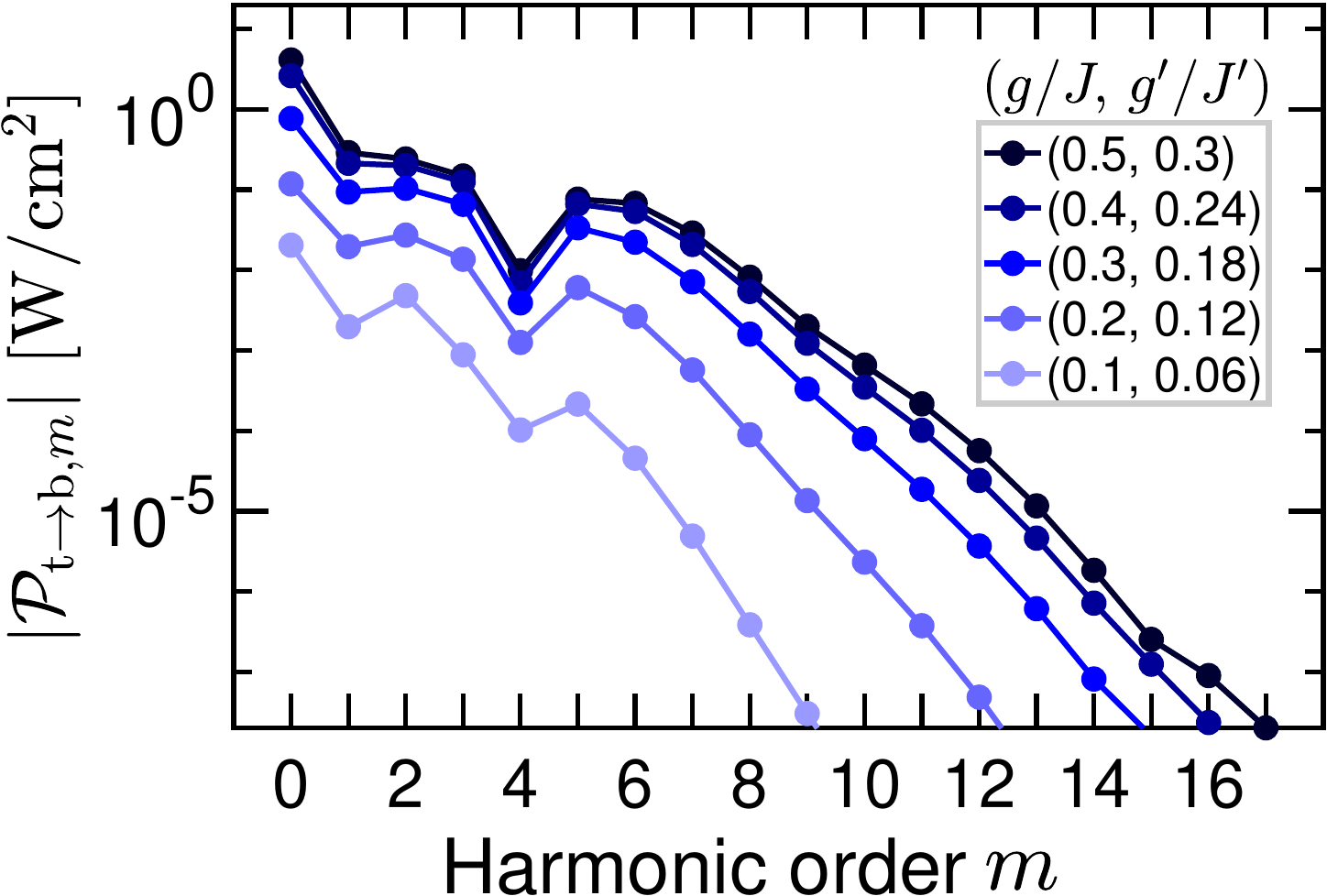}
		\caption{\textbf{Spin-phonon coupling effect on spin HHG}. Harmonic amplitudes and orders for different spin-phonon coupling strengths when $g > g'$ is fixed. As observed, harmonic amplitudes are enhanced by increasing spin-phonon coupling strengths, and the suppression of the $4$-th harmonic~(lower two-triplon band edge for $\Omega=\omega_0=\sqrt{2}J/4$) is visible. Other parameters are set to $\gamma_{\rm p}/\omega_0=0.05$, $\gamma_{\rm s}/J=0.015$, and $\mathcal{A}/ \gamma_{\rm p}=0.5$.}\label{fig3}
	\end{figure}

		As has been widely studied in the literature, HHG arises mainly from nonlinear interactions between the drive field and the target material~\cite{RevModPhys.72.545,Krausz_2016,ghimire2019high,li_attosecond_2020,yu2019high,goulielmakis_high_2022}. In the case of our model, the spin system is driven by the laser-driven phonon. Therefore, SPC would account for the nonlinear interaction between the driving and driven systems, leading to the HHG from the spin excitations. Figure~\ref{fig3} shows that the further we move away from the weak coupling regime, the more orders of harmonics appear in the frequency domain. Among all the coupling terms of phononic and triplon observables present in Eq.~\eqref{eq_36}, the coupling of phonon displacement $q(t)$ and the imaginary part of triplon pair creation $w_k(t)$~(spin excitation) have a major impact on the HHG, as turning off this convolution term would significantly decrease the number of harmonics in the spectral domain.
	
	Within the weak coupling regime~(small $g$ and $g'$), the excited phonon in the NESS can be perceived as a driven damped harmonic oscillator, thus, $q(t)=q_1\,\cos(\omega_0\,t)$ at $\Omega=\omega_{0}$. With this assumption and slowly varying component of triplon occupation in the NESS, we exactly solve Eqs.~\eqref{eq_36a},~\eqref{eq_36b}, and~\eqref{eq_36c} and find the analytical expression of $C_{\rm sp}(t) \equiv q(t) w_k(t)$ coupling term, given by
	\begin{subequations}
		\begin{align}
			&C_{\rm sp}(t)={}  \frac{\overline{\tilde{\xi}_k} q_1}{\chi_k} \Big[\mathcal{J}_1(\chi_k)+ \sum^\infty_{m=1} \widetilde{\mathcal{C}}_{k,m}\cos(m \,\omega_0 t+\tilde{\phi}_{k,m})\Big] \,, \label{eq_101}\\
			&\widetilde{\mathcal{C}}_{k,m}= [(m-1)\mathcal{J}_{m-1}(\chi_k)+(-1)^{m}(m+1)\mathcal{J}_{m+1}(\chi_k)] \notag\\&\times\sqrt{1+\Big[\underbrace{\mfrac{\delta[ (m-1)\mathcal{J}_{m-1}(\chi_k)-(-1)^{m}(m+1)\mathcal{J}_{m+1}(\chi_k)]}{\gamma_{\rm{s}}[(m-1)\mathcal{J}_{m-1}(\chi_k)+(-1)^{m}(m+1)\mathcal{J}_{m+1}(\chi_k)]}}_{=\,\tan(\tilde{\phi}_{k,m})}\Big]^2},
		\end{align}
	\end{subequations}where $q_1=2 \mathcal{A}/\gamma_{\rm p}$, $\chi_k={2 A_k} {q_1}/{\omega_0} $,  $\overline{\tilde{\xi}_k} = \frac{3\zeta_k \gamma_{\rm{s}}}{ 2 (\gamma_{\rm s}^2 -\zeta_k^2+ \delta^2)}$,
	$\zeta_k = \,\frac{B_k}{A_k}\, \omega_0 \, \mathcal{J}_1(\chi_k)$, $\delta = 2\varepsilon_k-\omega_{0}$ characterizes the phonon frequency deviation from the two-triplon band, and $\mathcal{J}_m(\chi_k)$ are the Bessel functions. At $\delta = 0$, we have the in-band scenario, i.e., $\omega_{0}=2\varepsilon_k$ and $\gamma_{\rm{s}}>|\zeta_k|$ must hold to achieve a positive triplon occupation $n_k(t)$ and convergence of the exponential terms of the observables at the long time limit. At $\delta \neq 0$, we have an out-of-band scenario (detuned), where $\gamma_{\rm{s}}>\sqrt{\zeta_k^2-\delta^2}$ must remain valid when $|\zeta_k|>|\delta|$. For the case of $|\zeta_k|<|\delta|$, no such conditions are needed. The full forms of $n_k(t), w_k(t), v_k(t)$, and calculation details are given in Appendix~\ref{apd}. For the strong coupling regime, the analytical expression is not straightforward.\begin{figure}[t]
		\centering
		\includegraphics[width=0.85\linewidth]{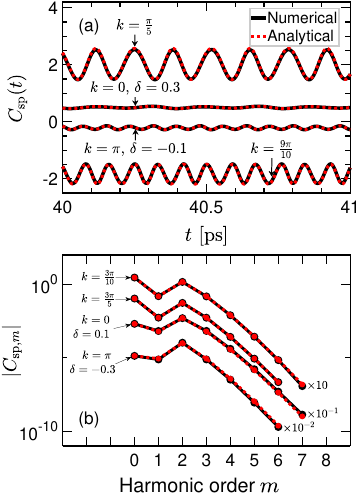}
		\caption{\textbf{Weak coupling theory for spin HHG}. The agreement between the numerical (black) and analytical (red) Eq.~\eqref{eq_101}, results of the $C_{\rm{sp}}(t)\equiv q(t)w_k(t)$ coupling term for different triplon modes $k$ at $\Omega=\omega_0$. Panel (a) displays the match between the time-evolution of $C_{\rm{sp}}(t)$ in the NESS, obtained by two methods, for the in-band~($\Omega = 2\varepsilon_k$) and out-of-band~($\Omega = 2\varepsilon_k-\delta$) scenario, respectively; the values of some of the modes are shifted vertically to avoid overlapping. Panel (b) illustrates the alignment between the HHG spectra of $|C_{{\rm{sp}},m}|$. Parameter sets are $\gamma_{\rm p}/\omega_0=0.05$,\ $\gamma_{\rm{s}}/J=0.02$,\ $\mathcal{A}/ \gamma_{\rm p}=0.5$, $g/J=0.2$, and $g'/J'=0.16$ in weak coupling regime.}\label{fig4}
	\end{figure}

	Figure~\ref{fig4}(a) demonstrates the agreement between numerical time-integration results and the analytical expressions stated above for different triplon modes $k$ in both in-band~($\delta = 0$) and out-of-band~($\delta \neq 0$) scenarios. The NESS time-evolutions of all these cases match perfectly. In Fig.~\ref{fig4}(b) we display the $|C_{\rm{sp,m}}|$ HHG spectra of all modes in both scenarios for which numerical and analytical HHG results denote a great level of consistency. We confirm that as the $C_{\rm sp}(t)$ coupling term is within the same order as the $\mathcal{P}_{{\rm t \to b}}(t)$, so it plays a crucial role in the harmonic contributions of the spin system. Note that the analytical expressions given above hold for small detuning $|\delta|$ values and by increasing $|\delta|$ the numerical and analytical results start to deviate from each other. 

	\subsection{Terahertz laser driving effects on spin HHG}\label{s3a1}
	
\begin{figure}[t]
		\centering
		\includegraphics[width=0.85\linewidth]{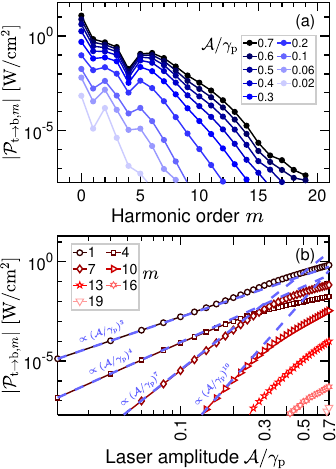}
		\caption{\textbf{Laser fluence effect on spin HHG}. (a) Spin HHG trends for a range of laser amplitudes $\mathcal{A}/\gamma_{\rm p}$ and (b) growth of the harmonic amplitudes for selected orders with increasing laser amplitude. According to panels (a) and (b), higher amplitudes and more orders of harmonics are expected when the laser amplitude is increased. The fitted dashed lines at low laser amplitude regimes correspond to power law scaling associated with the perturbative regime. For all cases, we set $\Omega=\omega_0=\sqrt{2}J/4$, $\gamma_{\rm p}/\omega_0=0.05$, $\gamma_{\rm s}/J=0.015$, $g/J=0.5$, and $g'/J'=0.3$.}
		\label{fig5}
	\end{figure}

	In this section, we aim to investigate the effect of drive parameters~(the laser amplitude $\mathcal{A}$ and drive frequency $\Omega$) on the spin HHG as the energy is pumped into the system by a continuous laser field. Figure~\ref{fig5}(a) portrays the HHG trends of $|\mathcal{P}_{{\rm t \to b},m}|$ for a range of laser amplitudes $\mathcal{A}/\gamma_{\rm p}$, while the rest of the parameters are fixed. As is apparent, increasing the laser amplitude amplifies the harmonic amplitudes and gives rise to higher-order harmonics. Moreover, Fig.~\ref{fig5}(b) demonstrates the growth of harmonic amplitudes for selected orders with respect to the laser amplitude $\mathcal{A}/\gamma_{\rm p}$, presented in a $\log\mbox{-}\log$ scale.
	We can recognize the linear behavior, which appears in the low laser amplitude regime, followed by a shift after a certain point to a different growth trend. The fitted dashed lines in Fig.~\ref{fig5}(b) suggest the power law scaling of harmonic amplitudes justifying a perturbative behavior. The harmonic amplitudes of order $=\{4,7,10\}$ clearly follow
	$|\mathcal{P}_{{\rm t \to b},m}|\propto (\mathcal{A}/\gamma_{\rm p})^m$, scaling with the order of process below $\mathcal{A}/\gamma_{\rm p}\approx0.2$. A similar behavior is observed in gas and solid HHG~\cite{ghimire_observation_2011,7052318,PhysRevB.99.184303,goulielmakis_high_2022} which is associated with a change from a perturbative regime to a non-perturbative regime. Although the $1$-st order harmonic follows a linear behavior within the perturbative regime, it scales with $(\mathcal{A}/\gamma_{\rm p})^{1+2}$. This indicates that for lower laser amplitudes the second harmonic is dominant over the first harmonic, see Fig.~\ref{fig5}(a), and by increasing the laser amplitude the first harmonic gradually rises and surpasses the second harmonic. This leads to a harmonic mixing and $|\mathcal{P}_{{\rm t \to b},1}|\propto (\mathcal{A}/\gamma_{\rm p})^3$. It is important to note that we cannot abruptly increase the laser intensity to obtain higher order as in real-world scenarios this leads to heating and eventually destruction of the material.
	
	Another phenomenon that we can notice from Fig.~\ref{fig5} is a crossing of amplitudes between the $4$-th and $7$-th harmonic. While one may expect the amplitudes of harmonic orders within a spectrum to decrease for the fixed set of parameters, the clearly visible harmonic mixing of Fig.~\ref{fig5} contradicts this statement. We observe that if the phonon frequency $\omega_0$ is almost equal to $1/m$ of either of the two-triplon band edges, i.e., $\omega_0 \simeq (1/m) 2 \varepsilon_k$ when $k=0$ or $k=\pi$, the amplitude of the $m$-th harmonic appears lower than the amplitude of harmonics immediately before and after, in the frequency domain. For some frequencies both conditions of $\omega_0 \simeq (1/m) 2 \varepsilon_{k=0}$, and $\omega_0 \simeq (1/m') 2 \varepsilon_{k=\pi}$ hold true; as an example, when $\omega_0=\sqrt{2}/4$, the frequency of the $4$-th harmonic is equal to $\sqrt{2}$, which lies on the lower edge of the two-triplon band, whilst the frequency of 7-th harmonic ($\approx 2.47$) is approximately equal to $\sqrt{6}$, which lies on the upper band edge of two-triplon. This suggests that when phonon frequency is located at the two-triplon band edges and the system is in the strong coupling regimes (large $g/J$ and $g'/J'$ strengths), the spin system strongly constrains the absorbed energy by the phonon sector that is delivered by the drive electric field source. The same effect extends itself even below the two-triplon band regime and clearly shows up in the HHG spectra of $|\mathcal{P}_{{\rm t \to b},m}|$. 
	
	In the next step, we investigate the specifications of spin HHG spectra within different drive frequency regimes. We identify four $\Omega/J$ regimes across the full frequency range of interest, i.e., $\Omega/J=0.2$ to $\Omega/J=3.0$, and the HHG trends of their showcased frequencies are illustrated in Fig.~\ref{fig6}. This range is of particular interest for covering the entire band, coinciding with the THz spectral range. Note that, to be able to work with frequencies below $0.2$, one needs to adjust selected spin and phonon damping parameters to ensure the condition $\gamma_{\rm p}>\gamma_{\rm s}$, as well as the NESS formation. The characteristics of each region are discussed below: \begin{figure}[t]
		\centering
		\includegraphics[width=0.85\linewidth]{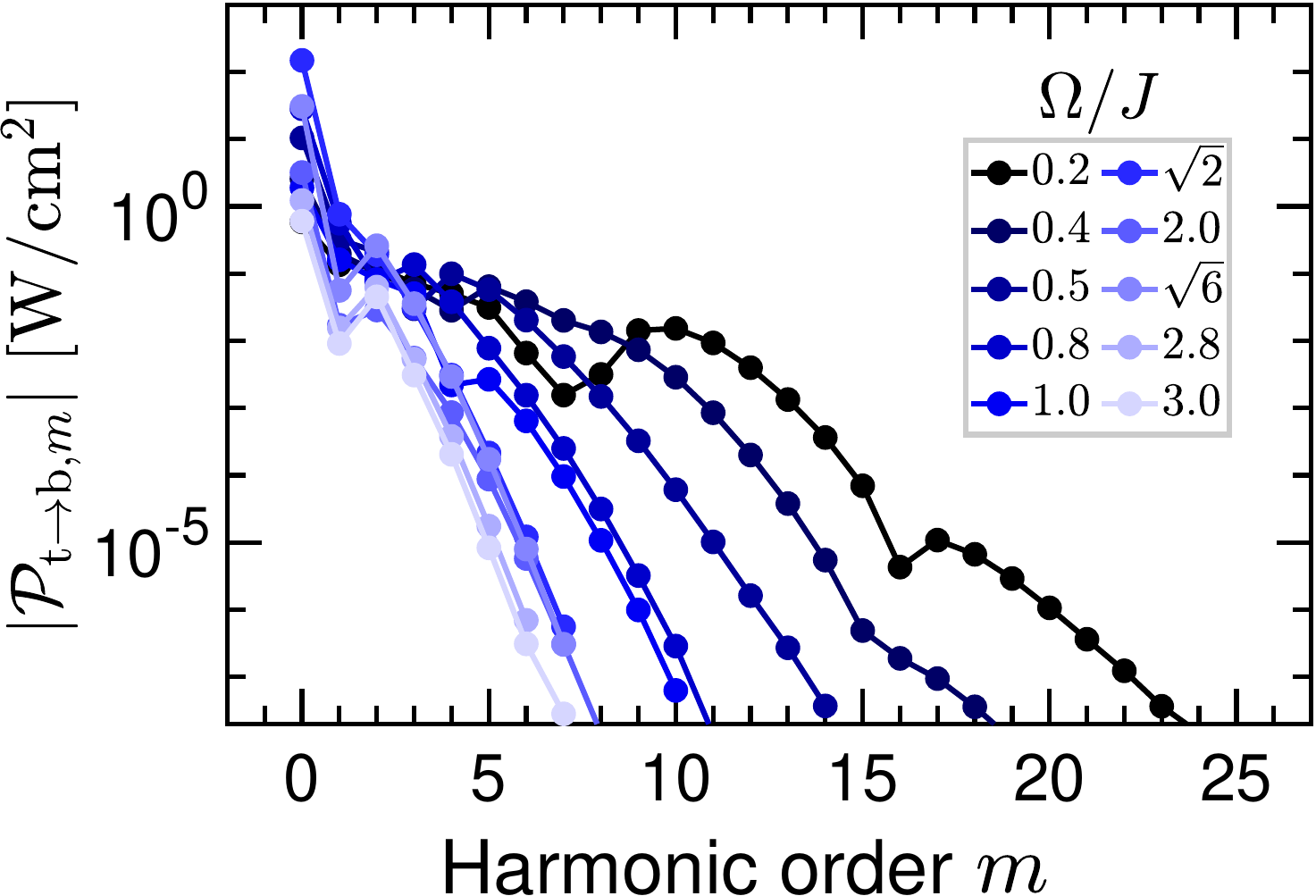}
		\caption{\textbf{Laser frequency effect on spin HHG}. Spin HHG spectra for different drive frequency regimes, namely, far below the two-triplon band ($\Omega/J=0.2,\ 0.4,~0.5$), below the band ($\Omega/J=0.8,~1.0$), in the band ($\Omega/J=\sqrt{2},~2.0,~\sqrt{6}$), and beyond 
 the band ($\Omega/J=2.8,~3.0$) are illustrated. In the low-frequency regime where the drive frequencies are located far below the two-triplon band, the highest orders of harmonics are generated from the spin system. For all cases, we fix $\Omega=\omega_0$ and the parameter sets are $\gamma_{\rm p}/\omega_0=0.06$, $\gamma_{\rm{s}}/J=0.011$, $\mathcal{A}/ \gamma_{\rm p}=0.5$, $g/J=0.5$, and $g'/J'=0.3$.
		} 
		\label{fig6}
	\end{figure}
	\begin{itemize}
		\item
		For $\Omega/J<0.5$ far below the two-triplon band, the number of harmonics is significantly large, for a fixed set of parameters. The two example frequencies, $\Omega/J=0.2$ and $\Omega/J=0.4$, illustrated in Fig.~\ref{fig6}, yield $23$ and $19$ harmonics, respectively. In this regime, we can pinpoint numerous frequencies that yield a suppression of certain harmonics in their spectrum, i.e., $\Omega/J$ is equal or close to $1/m$ of either or both two-triplon band edges. We previously discussed the suppression of the $4$-th harmonic for the case of $\Omega/J=\sqrt{2}/4$ in detail. Another instance is $\Omega/J=0.2$ which is almost 1/7~(1/16) of the lower two-triplon~(upper three-triplon) band edge, explaining the observed suppression at the 7-th~(16-th) harmonic in Fig.~\ref{fig6}.	
		\item 
		For $0.5<\Omega/J<\sqrt{2}$ below the two-triplon band, there is a notable decrease in the number of harmonics compared to the first region. As the illustrative cases of $\Omega/J=0.8$ and $\Omega/J=1.0$ reflect in Fig.~\ref{fig6}, $10$ orders of harmonics are observed in their HHG spectra. The one-triplon band, spanning from $\sqrt{2}/2$ to $\sqrt{6}/2$, falls within this region and we are expecting a slight increase in the harmonic amplitudes in the vicinity of the one-triplon band edges due to the resonant effects. On the other hand, moving toward the one-triplon mid-band, amplitude drops. This happens since for $k=\pi/2$ triplon mode, matrix elements of spin excitation become $\mathcal{S}_k = 0$ and $\mathcal{R}_k =1$, and eventually $A_k=g$ and $B_k=0$. This suggests a weak interaction between phonon and spin sectors for $k=\pi/2$ and therefore a lower HHG profile.
		As $\Omega/J=0.8$ is located close to the lower one-triplon band edge and $\Omega/J=1.0$ is located right at the one-triplon mid-band, harmonic amplitudes of $\Omega/J=0.8$ tend to be slightly larger than the harmonics of $\Omega/J=1.0$, except for the $2$-nd harmonic which are almost matching for both frequencies, meaning that $m \, \omega_0 \approx2\varepsilon_{k=0}$ is valid for $m=2$; thus the $2$-nd harmonic order appears slightly suppressed.      	
		\item
		For the two-triplon band, $\sqrt{2}\leq \Omega/J\leq \sqrt{6}$, there is a large increase in the zero-order harmonic amplitude or the average value close to band edges, even though they are slightly suppressed compared to the results of a weak SPC regime~\cite{PhysRevB.103.045132}. The $\Omega/J=\sqrt{2}$ illustrated in Fig.~\ref{fig6}, is located on the lower two-triplon band and attains the maximum average value, while the next maximum average value occurs at $\Omega/J=\sqrt{6}$ which is the upper two-triplon band. On the other hand, a decrease in harmonic amplitudes is anticipated in the mid-band where $\Omega/J=2.0$ is residing. The number of harmonics, however, remains more or less within the same order within this region around $7-10$ orders.		
		\item 
		Above the band, $\sqrt{6}<\Omega/J<3.0$, one would not expect an interesting behavior since we will not encounter band edges, therefore, we expect a monotonic decrease in the amplitudes of harmonics and a decrease in harmonic orders. The $\Omega/J=2.8$ and $\Omega/J=3.0$ in Fig.~\ref{fig6} exemplify the behavior of this regime. One thing to note is that the second harmonic order will be dominant over the first harmonic throughout this region. This aligns with the behavior in the weak regime, where the second harmonic was dominant over the first harmonic almost for the whole region. According to Eq.~\eqref{eq_36a} $n_k(t)$ is dominated by the $q(t)w_k(t)$ coupling term and as discussed in the Appendix.~\ref{sec:offresonantetc}, there is predominately first harmonic with frequency of $\omega_0$ in $q(t)$. The same oscillation exists in $w_k(t)$, thus making the 2$\omega_0$ frequency dominant in $n_k(t)$, and eventually, the same behavior is expected from $\mathcal{P}_{{\rm t \to b}}(t)$.
	\end{itemize}
	To recap, the overall behavior of the spin HHG spectra is a decrease in the orders of harmonics as the drive frequency $\Omega/J$ approaches the two-triplon band (spin band). The amplitudes of high-order harmonics will decrease even more compared to preceding orders. 
	
	\subsection{Experimental perspective}
	
	As the leading magnetic order $J$ of our target material CuGeO$_3$, is approximately 2.4 THz, in case the objective is to generate maximum orders of spin harmonics, as discussed in Sec.~\ref{s3a1}, it is more favorable to sample drive frequencies far below the two-triplon band, e.g., $\Omega/J<0.5$ means drive frequency less than $1.2$ THz. This range is deemed accessible by current continuous wave THz laser sources that are utilized in various applications such as imaging, spectroscopy, etc.~\cite{Biasco2018, Khalatpour2021, Liebermeister2021, COSTA2021108904, WAN2020105859, Yang2023}. Additionally, to maintain a dimerized CuGeO$_3$, the temperature of the experiments must be set to a few Kelvin~\cite{PhysRevLett.70.3651}. Using Eq.~\eqref{eq:power} in the NESS, we can gauge the time for which the system maintains the NESS necessary for observing harmonics. This time, denoted as $t_{\rm HHG}$, can be estimated as $t_{\rm HHG} = m \int_0^T C(T) \, d\,T/|\mathcal{P}_{{\rm t \to b},0}|$. 
 
    Here, the sample interfaces with a cold sink characterized by a temperature range of $T \approx 2-5$ K, a mass range of $m \approx 2-5$ g, and a small heat capacity $C(T) \propto T$ j.g$^{-1}$.K$^{-2}$. With a spin emitted power of approximately 2 W/cm$^2$, harmonic generation can be observed for durations up to $t_{\rm HHG} \approx 2$ microseconds, a timescale well within the picosecond NESS duration simulated in our study. Once the setup is prepared, and the NESS is achieved, the spin HHG could be measured from the energy outflow of magnetic excitations to the environment with the help of magnetization in the presence of a uniform Zeeman magnetic field~\cite{10.1063/1.5054116,goulielmakis_high_2022,PhysRevLett.123.197204,Tashiro2015,Chen2021}.
	

	\section{Conclusions}\label{s4}
	
	Steady states are not only essential for elucidating fundamental concepts in condensed matter physics but also hold significance in the development of quantum technologies such as quantum sensing, computation, and information processing. Here, we have focused on the steady states of a driven-dissipative quantum magnet to uncover unique high-harmonic patterns originating from magnetic excitations, diverging from the predominant focus on electronic systems in high-harmonic generation (HHG) studies. We employ a magnetophononic framework such that the lattice vibrations, driven by the laser's electric field, stimulate the spin system, obviating the need for a strong magnetic field. We have used a terahertz continuous-wave laser field to pump energy into the system, allowing the occurrence of non-equilibrium steady states and laying the foundation for HHG. This allows us to govern the system dynamics, offering a more realistic scenario leading to HHG via spin-phonon coupling. 
	
	The strong spin-phonon couplings, entailing the coupling of phonons to both intradimer and interdimer spin-spin interactions, enhance the mutual feedback effect between the two sectors. This hybridization effect fosters nonlinear interactions, resulting in spin HHG. Notably, among all coupling terms in the dynamics of observables, the coupling of lattice displacement to the spin excitation plays a vital role in spin HHG emergence. To address this, we have proposed analytical expressions for this spin-phonon coupling term within the weak regime, validating our numerical findings. By scanning the HHG spectrum within the permissible range of the drive electric field amplitude~(following the dimerization phase of our spin model at low temperatures), we observed an increase in both the number of harmonic orders and the amplitudes of the harmonics with larger laser amplitudes. We also found that drive electric field frequencies far from the dispersion energy of the two-triplon band generate the highest orders and harmonic amplitudes, making this range suitable for setups requiring an optimal number of high harmonics. An increase in damping coefficients only marginally enhances harmonic amplitudes.
	
	With all that said, we believe that integrating HHG into quantum spin systems for near-future technologies will expand the horizons of current HHG applications, including spectroscopy, coherent imaging, and ultrafast dynamics. Additionally, it offers exciting opportunities to use the unique features and applications of the terahertz spectral range across diverse fields, particularly in spintronics. 
	
	\section*{Acknowledgments}
	M.Y. greatly thanks G\"otz S. Uhrig and Bruce Normand for helpful discussions. This work was performed with support from the National Science Foundation (NSF) through award numbers MPS-2228725 and DMR-1945529 (M.Y. and M.K.). Part of this work was performed at the Aspen Center for Physics, which is supported by NSF grant No. PHY-1607611. Funded by the European Union (ERC, QuSimCtrl, 101113633). Views and opinions expressed are however those of the authors only and do not necessarily reflect those of the European Union or the European Research Council Executive Agency. Neither the European Union nor the granting authority can be held responsible for them. The Flatiron Institute is a division of the Simons Foundation.
	
	\appendix
        \section{Dimerization effect on spin HHG}\label{apa}\begin{figure}[t]
		\centering
		\includegraphics[width=0.85\linewidth]{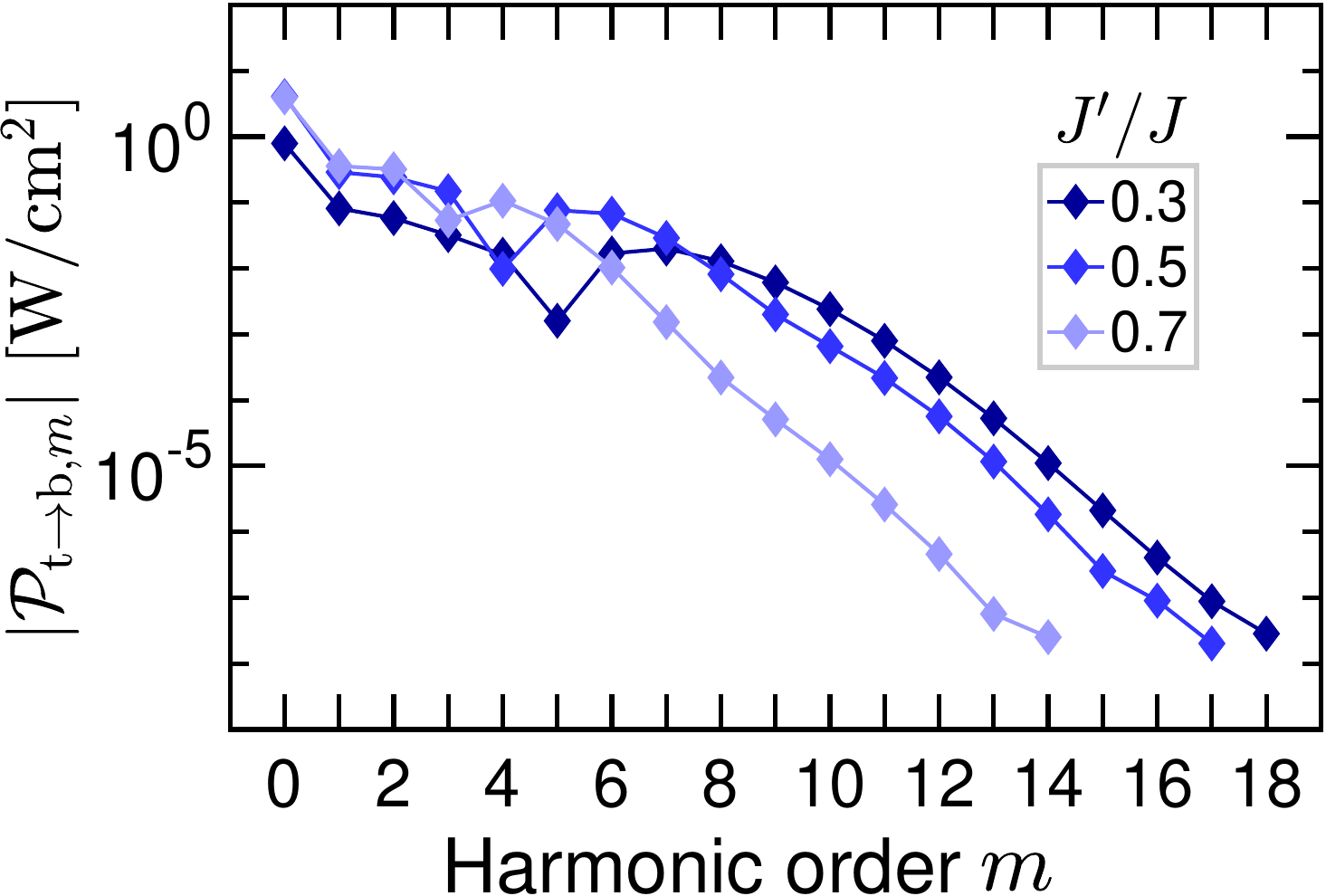}
		\caption{\textbf{Dimerization effect on spin HHG}. When the dimerization factor $J'/J$ decreases, we observe a slightly larger number of spin harmonics in the HHG profile. This indicates that a narrower spin band tends to favor the generation of more harmonics. Parameters are $\Omega=\omega_0=\sqrt{2}J/4$, $\gamma_{\rm p}/\omega_0=0.05$, $\gamma_{\rm{s}}/J=0.015$, $\mathcal{A}/\gamma_{\rm p}=0.5$, $g/J=0.5$, and $g'/J'=0.3$. As discussed in the main text, it is worth noting that the ultralow frequency regime tends to yield the highest number of harmonics.}\label{fig7_new}
	\end{figure}

In this section, to investigate how narrowing or broadening the spin band influences spin HHG, we have examined two additional values of the dimerization factor $J'/J$, namely $J'/J= 0.3$ and $J'/J= 0.7$, which differ from the selected value of $J'/J = 0.5$ in the main text. Figure~\ref{fig7_new} shows that as the dimerization factor $J'/J$ decreases, the number of spin harmonics increases in the HHG profile. This trend, in turn, suggests that the narrowing of the spin band appears to facilitate the generation of additional harmonics.
 
	\section{Dissipation effects on spin HHG}\label{apb}
	\begin{figure}[t]
		\centering
		\includegraphics[width=0.85\linewidth]{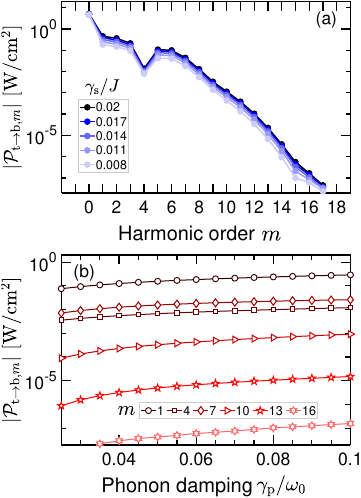}
		\caption{\textbf{Dissipation effect on spin HHG}. (a) Spin HHG trends by varying spin damping $\gamma_{\rm{s}}/J$ at $\gamma_{\rm p}/\omega_0=0.06$. (b) Growth of the amplitudes by varying phonon damping $\gamma_{\rm p}/\omega_0$ at $\gamma_{\rm{s}}/J=0.008$. Parameters are $\Omega=\omega_0=\sqrt{2}J/4$,  $\mathcal{A}/\gamma_{\rm p}=0.5$, $g/J=0.5$, and $g'/J'=0.3$.}\label{fig7}
	\end{figure}
	Considering that dissipation effects are introduced to our open quantum system by Lindblad formalism, we seek to know whether the damping parameters affect the spin HHG. Figure~\ref{fig7}(a) demonstrates how the harmonic orders of $|\mathcal{P}_{{\rm t \to b},m}|$ evolve by increasing the spin damping $\gamma_{\rm{s}}/J$, while the rest of the parameters are fixed. Within the applicable range of $\gamma_{\rm{s}}/J$ for the selected parameter set, increasing $\gamma_{\rm{s}}/J$ will slightly increase the harmonic amplitudes of $|\mathcal{P}_{{\rm t \to b},m}|$. Although, according to Fig.~\ref{fig7}(a) opting for a larger $\gamma_{\rm{s}}/J$ seems to be advantageous, configuring $\gamma_{\rm{s}}/J$ is constrained by a couple of factors; one point is that the spin excitations are generally weakly coupled to the phononic bath, leading to a small $\gamma_{\rm{s}}/J$ value for the system. Another factor to keep in mind is that attaining a NESS in the long-time limit of the observable is crucial, which appears to be violated below $\gamma_{\rm{s}}/J=0.008$, for the parameter set of Fig.~\ref{fig7}(a), thus they are excluded from the range of study. The permitted range of $\gamma_{\rm{s}}/J$ might be broader for larger $\Omega/J$, when the system is set to the same $\gamma_{\rm p}/\omega_0$, however, according to our analysis presented in the previous sections, staying within the low $\Omega/J$ regime would generally result in high-orders of harmonics and larger harmonic amplitudes in the HHG spectrum of $|\mathcal{P}_{{\rm t \to b},m}|$.\begin{figure}[b]
	\centering
	\includegraphics[width=0.85\linewidth]{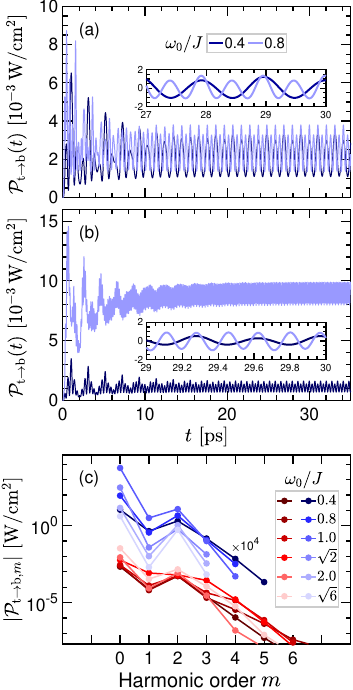}
	\caption{\textbf{Weak off-resonance spin HHG}. The spin HHG spectra for $\Omega \neq \omega_0$, i.e., off-resonance settings. Panel (a) and (b) confirm the NESS formation for (a) $\Omega<\omega_0$ ($\Omega=0.5\, \omega_0$) and (b) $\Omega>\omega_0$ ($\Omega=1.5 \,\omega_0$) cases, respectively. The inset panels are demonstrations of signal patterns in the NESS, and for comparison, the average values are deducted from the original values. Panel (c) shows the HHG spectra for cases $\Omega=0.5\, \omega_0$ (red), and $\Omega=1.5\, \omega_0$ (blue) over a range of $\omega_0/J$ values. As visible from these plots, orders of harmonics have decreased in all cases compared to the resonance scenario. Parameter sets are $\gamma_{\rm p}/\omega_0=0.06$, $\gamma_{\rm{s}}/J=0.011$, $\mathcal{A}/ \gamma_{\rm p}=0.5$, $g/J=0.5$, and $g'/J'=0.3$.}\label{fig8}
\end{figure}  
\begin{figure}
    \centering
    \includegraphics[width=0.85\linewidth]{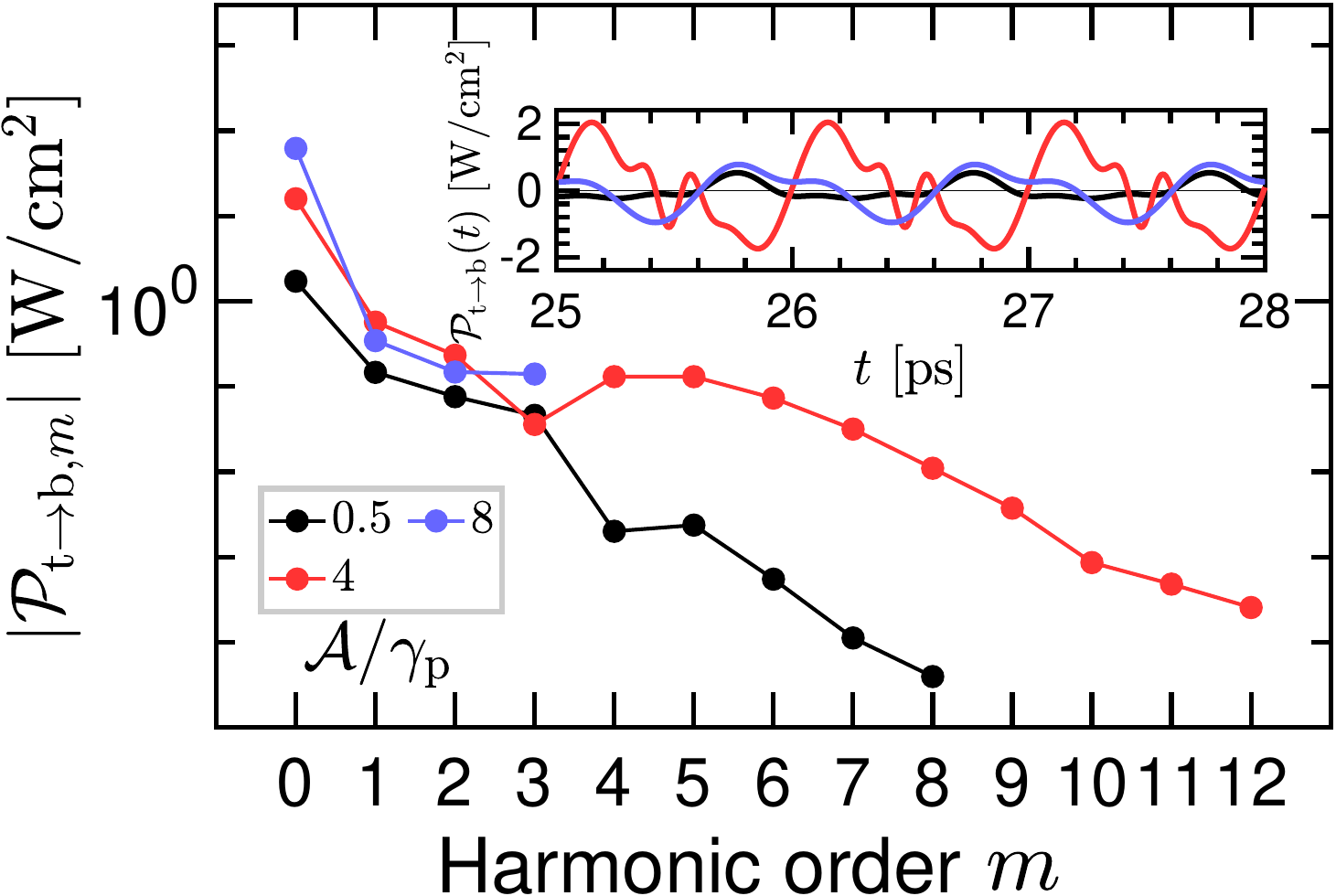}
    \caption{\textbf{Strong off-resonance spin HHG}. We show red- and blue-detuned NESS with \{$\Omega=0.5 \,\omega_0$ and $\mathcal{A}/\gamma_{\rm p} = 4$\} and \{$\Omega=1.5 \,\omega_0$ and $\mathcal{A}/\gamma_{\rm p} = 8$\}, respectively, and contrast against resonant excitation with \{$\Omega= \,\omega_0$ and $\mathcal{A}/\gamma_{\rm p} = 0.5$\}. The inset panel shows the time-trace of the three latest periods of NESS with static averages subtracted. 
        Parameters used in this plot: $\omega_0/J=1$, $\gamma_{\rm p}/\omega_0= 0.05$, $\gamma_{\rm s}/J=0.01$, $g /J =0.5$, and $g'/J'=0.3$.} 
    \label{fig10}
\end{figure}
	
	Next, we address the effect of the phonon damping $\gamma_{\rm p}/\omega_0$ on the spin HHG spectra. Note that, on the occasion of tuning the proper parameter set to achieve the optimum outcome of the spin HHG spectra, one needs to consider that a driven phonon is weakly coupled to bath while maintaining $\gamma_{\rm p}> \gamma_{\rm{s}}$ status. We let the $\gamma_{\rm p}$ scale with phonon energy to monitor the damped energy relative to the phonon energy. Figure~\ref{fig7}(b) demonstrates the evolution of harmonic amplitudes of $|\mathcal{P}_{{\rm t \to b},m}|$ by increasing $\gamma_{\rm p}/\omega_0$, while the rest of the adjustable parameters are kept constant. In this arrangement, an increase in $\gamma_{\rm p}/\omega_0$ (with constant $\omega_0/J$) is in conjugate with an increase in laser amplitude through $\mathcal{A}/\gamma_{\rm p}$. Therefore, as seen in Fig.~\ref{fig5} we are expecting a form of increase in harmonic amplitudes with increasing $\gamma_{\rm p}/\omega_0$. However, within the accessible range of $\gamma_{\rm p}/\omega_0$, as discussed earlier, it may not be an effective factor in increasing harmonic orders of the spin HHG.

	\section{Off-resonant laser drive and stability of NESS}\label{sec:offresonantetc}
	Results presented in the main text focus on the resonant case with laser frequency $\Omega$ matches that of phonon $\Omega=\omega_0$.  Both are taken to be of order spin-interactions, i.e., the phonons being driven are themselves the result of the spin-dimerization. The purpose of this Appendix is to consider off-resonant driving, where much stronger laser amplitude may be tolerated. The upshot of our simulations (see below) is that blue- and red- detuning produces qualitatively different excitation patterns, with only red-detuned driving resulting in strong plateau-like HHG profile.
 We have not pursued a complete understanding of this dichotomy, only noting in passing its similarity to celebrated HHG in attosecond pulse generation, where lower frequency intense driving was key to entering the non-perturbative regime\cite{attosecondHHGPhysicsToday}. 

 We start the exploration of off-resonant by scanning the parameter space at fixed (relatively weak) amplitude of the drive (Fig. \ref{fig8}) and then proceed to much stronger driving amplitude (Fig. \ref{fig10}). As per usual convention we examine red- and blue-detuned cases, for which $\Omega=0.5 \,\omega_0$ and  $\Omega=1.5 \,\omega_0$, respectively.
Panels (a) and (b) of Fig.~\ref{fig8} display that the NESS is established in both off-resonance cases. However, judging by the oscillations of the inset panels, we can presume that there are not so many harmonic complexities compared to the inset panels of Fig.~\ref{fig2} in which the resonance scenario is represented. Accordingly, the HHG plots in Fig.~\ref{fig8}(c), exhibit $5$-$6$ orders of harmonic overall for the sampled frequencies, where we observed up to $23$ orders of harmonics for resonance under the same set of parameters. Therefore, we identify a large decrease in the orders of harmonics for the off-resonance scenario compared to the resonance. The relatively small values used here imply a significantly larger (order of magnitude!) driving amplitude required to achieve a similar degree of displacement.
The results for strong off-resonant driving (Fig.~\ref{fig10}) reveal a rich pattern of excitation, with appreciable rectification, i.e., average DC displacement of both phonon and triplons (these were subtracted in plots), alongside diverse patterns of HHG. We have also observed clear phase synchronization among the harmonics (\cite{attosecondHHGPhysicsToday} in the red-detuned case which is absent in the blue-detuned cases. The resonant case also exhibits synchronization with $\pi/2$ phase shift with respect to the laser and $\pi$ shifts between odd/even harmonics. We plan to explore these structures in future work.\begin{figure}[t]
			\centering
			\includegraphics[width=0.85\linewidth]{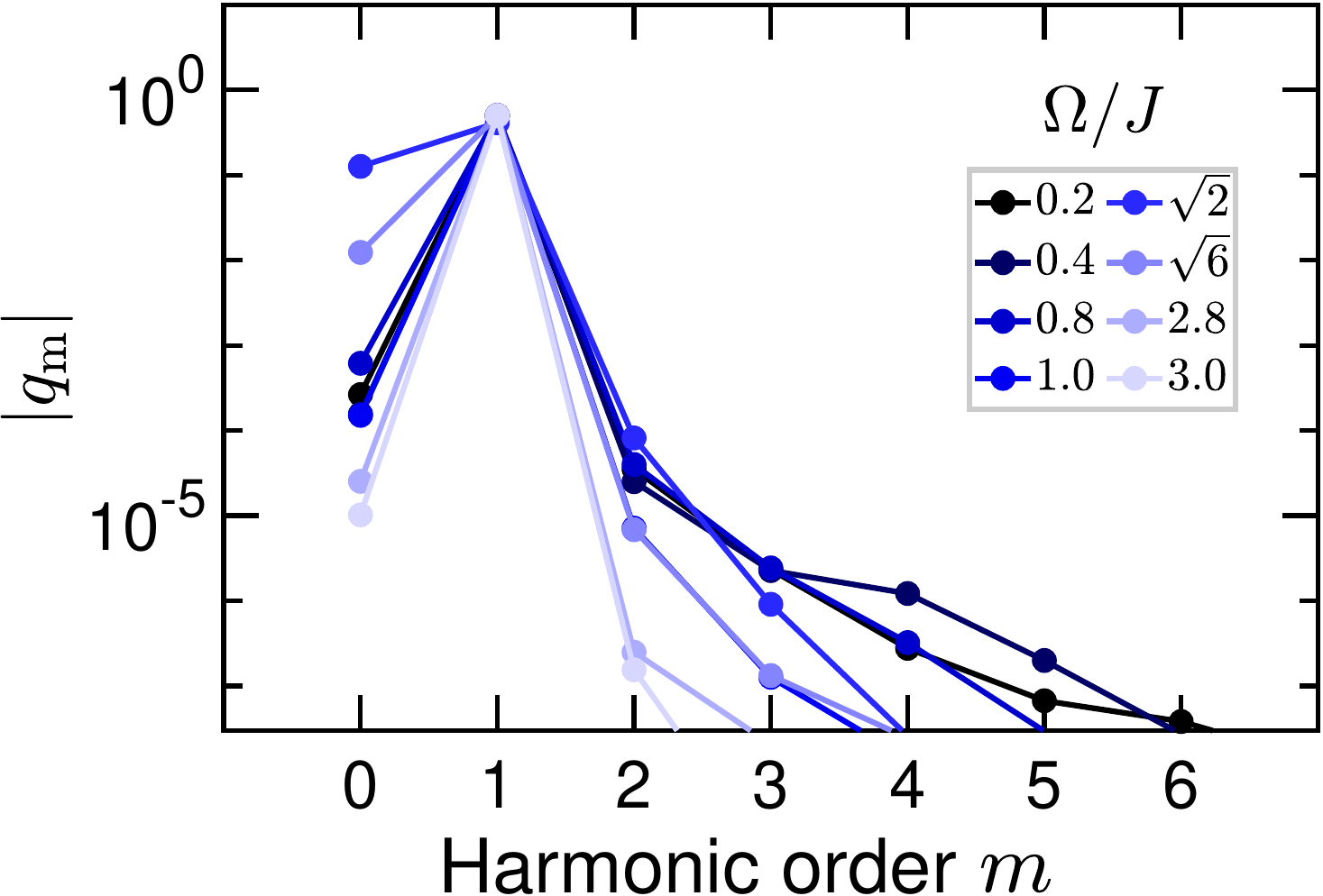}	
			\caption{\textbf{Phonon displacement harmonics}. The HHG profile of $|q_{\rm{m}}|$ (harmonic decomposition of phonon displacement $q(t)$) in weak coupling regime, obtained by numerical method. As observed in the plots, there are sharp peaks located at the first harmonic with the amplitude of $\mathcal{A}/ \gamma_{\rm p}=0.5$, dominant over the rest of the harmonic orders. The parameter sets are $\Omega=\omega_0$, $\gamma_{\rm p}/\omega_0=0.06$, $\gamma_{\rm{s}}/J=0.011$, $g/J=0.2$, and $g'/J'=0.16$.}\label{fig9}
		\end{figure}
  
Finally, we close with a discussion of two \emph{types} of limitations of our mean-field-like modeling of the problem: First, and most transparent physically, we require a small degree of excitation of both triplons and phonons, to maintain them both sufficiently dilute to justify ignoring other nonlinearities. For the triplon sector, we target the triplon number $t^\dagger t\lesssim 0.1$, while for the phonon mode, we borrow from the well-known Lindemann criterion for thermal melting, whereby root-mean-square displacement of a lattice point is bounded by a tenth of a lattice spacing, approximately. Importantly, for a typical Einstein phonon, the stiffness constant is itself set by the local geometry of the unit cell (i.e. lattice spacing), and so the Lindemann criterion can be recast \cite{yarmohammadi2022non} to a maximal dimensionless value of the average phonon number at the transition of about 10.  In the units where $\hbar=J=m_{\rm atomic}=1$ this implies $\langle {\bf q}^2\rangle\approx 10$. We can define and use $|q(t)|\leq 3$ as the dynamical analog \cite{mankowsky2017dynamical} of the Lindemann criterion \cite{vopson2020generalized}, even when the motion is not thermal, as happens to be the case here. So, while the phonon excitation we consider here displays comparatively weaker HHG than triplons, we nevertheless monitor the overall amplitude which remains well within the dynamic Lindemann bound. 

The second, and more mathematical, limitation originates with the strictly classical nature of the dynamical system we obtain after truncating the quantum equations of motion. To be precise, while the underlying fully quantum problem obeys the Floquet theorem, the mean-field approximation results in a classical nonlinear dynamical system for which we are not aware of an analogously powerful guarantee of periodic NESS induced by periodic driving. We do expect, under some favorable conditions, that the system will settle into a limit cycle \cite{chan2015limit}(or a more elaborate type of periodic attractor), but there is no reason for such attractors to be unique, nor to completely exhaust the possibilities; indeed, we have empirically observed the breakdown of NESS, roughly when the amplitude of the phonon displacement exceeds 1 or so (in our dimensionless units), i.e., just before the dynamical Lindemann criterion is reached. 

  \onecolumngrid
		{
		\section{Weak spin-phonon coupling theory for spin HHG with phonon driven at $\Omega = \omega_0$}\label{apd}
		
		In this appendix, we will propose analytical expressions regarding the generated harmonics from equations of motions Eqs.~\eqref{eq_29} and \eqref{eq_36}, and demonstrate the importance of the $q(t) w_k(t)$ coupling term in developing the high-order harmonics. For this purpose, we will adopt certain simplifications to reduce the complexities while elaborating more on the groundwork of the process.
		
		We start with a cosinusoidal signal in the NESS, oscillating with frequency $\Omega = \omega_0$, forming the temporal state of the phonon displacement $q(t)$, given by: 
		\begin{equation}\label{eq_42}
			q(t)=q_1\,\cos(\omega_0\,t)\, .
		\end{equation}
		This is considered a valid assumption since based on the HHG spectrum of $q(t)$, as denoted in Fig.~\ref{fig9}, a sharp peak is located on the first harmonic which is dominant over the rest of the harmonics.
		On the other hand, within a weak coupling regime of SPC, the excited phonon is perceived as a driven damped harmonic oscillator, therefore, the amplitude of its oscillation is roughly equal to $2\mathcal{A}/\gamma_{\rm p}$ and thereby $q_1=2\mathcal{A}/\gamma_{\rm p}$.
		
		Through the use of Eqs.~\eqref{eq_36b} and \eqref{eq_36b}, and knowing that $v_k(t) = {\rm Re}\, z_k(t)$ and $w_k(t) = {\rm Im}\, z_k(t)$, we obtain
		\begin{equation}
			\dot z_k(t) =  \big[ 2\, i \big(\varepsilon_k + A_k \,q(t)\big)-\gamma_{\rm{s}}\big]\, z_k(t) +\, 2\, i \,B_k\, q(t)\, \big(n_k(t) + {\textstyle \frac{3}{2}}\big) .\label{eq_46}		
		\end{equation}
		By defining $\eta_k(t) =  {} 2\varepsilon_k \, t + \chi_k \sin(\omega_0 t)$ and $f_k(t) =  {} 2\, B_k \,q(t)\,\big(n_k(t) + {\textstyle \frac{3}{2}}\big)$,	where $\chi_k={2 A_k} {q_1}/{\omega_0} $, the solution of above equation becomes
  \onecolumngrid
		{ 
		\begin{equation}
			z_k(t)=  \, 2i \,B_k \,q_1 \, e^{i \eta_k(t)
				-\gamma_{\rm{s}} t} \underbrace{\int_{0}^{t} \Big[\cos{(\omega_0 t')}\, \big(n_k(t') + {\textstyle \frac{3}{2}}\big)\, e^{- 2i\varepsilon_k t' -2i \chi_k \sin(\omega_0 t')
					+\gamma_{\rm{s}} t'} dt'\Big]}_{\rm I} \, .\label{eq_51}	
		\end{equation}
		
		Moving onward, we focus only on the slowly varying component of $n_k(t)$ and not the fast-oscillating terms, therefore we are allowed to average over one period, $T=2\pi/\omega_0$. For now, we work within the in-band region, meaning that $\Omega=\omega_0=2\varepsilon_k$. We will extend the same procedure to the out-of-band (detuned) scenario where the phonon frequency deviates from the two-triplon band, i.e, $\Omega=\omega_0=2\varepsilon_k-\delta$ which will be presented later in the same section. These simplifications allow us to write the integral term of Eq.~\eqref{eq_51} as follows
		\begin{equation}
			{\rm I}= 
			\int_{0}^{t}  \Big[{ \frac{1}{T}} \int_{0}^{T} \cos{(\omega_0 t'')}\,\, {} e^{-i\omega_0 \, t'' -\, 2i \, \chi_k \, \sin(\omega_0 t'')}\,
			dt''\Big]\big(n_k(t') + {\textstyle \frac{3}{2}}\big)\, e^{\gamma_{\rm{s}} t'} \,dt' \, .\label{eq_52}
		\end{equation}
		With the aid of Jacobi-Anger expansion:
		\begin{subequations}
			\begin{align}
				\cos\big(\chi_k \sin(\omega_0 t)\big)=\, & \mathcal{J}_0(\chi_k)+2 \sum^\infty_{m=1} \mathcal{J}_{2m}(\chi_k) \cos\big(2 m\, \omega_0 t\big) ,\label{eq_43}\\
				\sin\big(\chi_k \sin(\omega_0 t)\big)=\, & 2 \sum^\infty_{m=1} \mathcal{J}_{2m-1}(\chi_k) \sin\big((2 m-1) \, \omega_0 t\big)\, ,\label{eq_44}
			\end{align}
		\end{subequations}
		where $\mathcal{J}_m(\chi_k)$ are the Bessel functions of the first kind. Along with the recurrence relation $\frac{2m}{\chi_k}	\mathcal{J}_m(\chi_k)=\mathcal{J}_{m-1}(\chi_k)+\mathcal{J}_{m+1}(\chi_k)$ and by performing series of calculations, Eq.~\eqref{eq_51} yields
		\begin{equation}
			z_k(t) =  i\,\underbrace{\frac{B_k}{A_k} \, \mathcal{J}_1(\chi_k)\, \omega_0}_{ \zeta_k} \, e^{i \eta_k(t)
				-\gamma_{\rm{s}} t} 		\int_{0}^{t}   \big(n_k(t') + {\textstyle \frac{3}{2}}\big)\, e^{\gamma_{\rm{s}} t'}\,dt'\, .	 \label{eq_56}	
		\end{equation}

		Afterward, we take Eq.~\eqref{eq_36a} along with Eq.~\eqref{eq_42} and $w_k(t)= -(i/2)\big(z_k(t)-z_k^*(t)\big)$, to obtain
		\begin{equation}
			\dot n_k(t)+\gamma_{\rm{s}} \, n_k(t) =\,-i\,B_k\, q_1 \cos{(\omega_0 t)}\,\big(z_k(t)-z_k^*(t)\big)\, .
		\end{equation}
		Next, using Eq.~\eqref{eq_56}, we average over one period and obtain
		\begin{equation}
			n_k(t)\,  =\zeta_k^2 e^{-\gamma_{\rm{s}} t}\int_{0}^{t} \int_{0}^{t'}   \big(n_k(t'') + {\textstyle \frac{3}{2}}\big)\, e^{\gamma_{\rm{s}} t''} \,dt''\,\,dt' \, .
			\label{eq_59}
		\end{equation}
		After taking the derivative of the above equation twice, and considering the appropriate initial conditions, i.e., $n_k(0)=0$, and $d\,n_k(0)/dt=0$, the solution is
		\begin{equation}
			\begin{aligned}
				\label{eq_40}
				n_k(t) = \frac{3\, \zeta_k ^2}{2 (\gamma_{\rm s}^2 - \zeta_k ^2)} \Big[&1+\frac{1}{2}(\frac{\gamma_{\rm s}}{\zeta_k}-1)e^{-(\gamma_{\rm s}+\zeta_k) \,t}-\frac{1}{2}(\frac{\gamma_{\rm s}}{\zeta_k}+1)e^{-(\gamma_{\rm s}-\zeta_k) \,t} \Big]  \, .
			\end{aligned}
		\end{equation}
		The $n_k(t)$ observable denotes the triplon occupation of $k$-th mode and it cannot have a negative value. Moreover, $n_k(t)$ should converge to a constant value at $t\to\infty$ limit, to ensure the NESS establishment. Thus, $\gamma_{\rm s}>|\zeta_k|$ condition is the requirement of Eq.~\eqref{eq_40}. The long time behavior of this equation in the NESS yields $\lim_{t \to \infty} n_k(t) = \frac{3\, \zeta_k ^2}{2 (\gamma_{\rm s}^2 - \zeta_k ^2)}$, representing the average value of $n_k(t)$ as well as the amplitude of $0$-th order harmonic.

		We utilize Eq.~\eqref{eq_40} along with Eq.~\eqref{eq_42} to find $v_k(t)$ and $w_k(t)$ such that construction of additional orders of harmonics become visible. We insert it into the Eq.\eqref{eq_56} to obtain the full form of $z_k(t)$ in the NESS~($t \to \infty$):
		\begin{equation}
			\begin{aligned}
				z_k(t)&=  \underbrace{- \overline{\xi_k} \sin\big(\eta_k(t)\big)}_{v_k(t)} + i \underbrace{\overline{\xi_k}\cos\big(\eta_k(t)\big)}_{w_k(t)}\, ,\label{eq_60}
			\end{aligned}		
		\end{equation}where $\overline{\xi_k} = \frac{3\,\zeta_k \gamma_{\rm s}}{ 2(\gamma_{\rm s}^2 - \zeta_k ^2)}$.
		The $\cos\big(\eta_k(t)\big)$ term of $w_k(t)$, and $\sin\big(\eta_k(t)\big)$ of $v_k(t)$ mainly contribute to higher-order harmonics. To have a clearer picture, we decompose $\cos\big(\eta_k(t)\big)$ term of $w_k(t)$ into $\cos\big(\eta_k(t)\big)=\, \cos(\omega_0 t)\,\cos\big(\chi_k \sin(\omega_0 t)\big)-\sin(\omega_0 t)\,\sin\big(\chi_k \sin(\omega_0 t)\big)$. Next, by using the Eqs.~\eqref{eq_43} and~\eqref{eq_44}, the two terms of the expression above are expanded as
		\begin{subequations}
			\begin{align}
				\cos(\omega_0 t)\,\cos\big[\chi_k \sin(\omega_0 t)\big]= {} &
				\sum^\infty_{m=1} \Big[\mathcal{J}_{2m-2}(\chi_k) +  \mathcal{J}_{2m}(\chi_k)\Big]\cos\big[(2m-1) \,\omega_0 t\big]\, , \label{eq_63}\\
				\sin(\omega_0 t)\,\sin\big[\chi_k \sin(\omega_0 t)\big]= {} &\mathcal{J}_1(\chi_k)+\sum^\infty_{m=1} \Big[-\mathcal{J}_{2m-1}(\chi_k) +  \mathcal{J}_{2m+1}(\chi_k)\Big]\cos\big[(2m) \,\omega_0 t\big]\, .\label{eq_64}
			\end{align}
		\end{subequations}
		As mentioned in the main text, the primary source of spin HHG is the coupling between the driven phonon sector and the triplon sector, in which the $q(t)w_k(t)$ coupling has a substantial contribution. Thus, we have
		\begin{equation}
			\begin{aligned}
				q(t) w_k(t)={} & \frac{\overline{\xi_k} q_1}{\chi_k}\Big(\mathcal{J}_1(\chi_k)+
				\, \sum^\infty_{m=1} \Big[(m-1)\mathcal{J}_{m-1}(\chi_k) +(-1)^{m}(m+1)\mathcal{J}_{m+1}(\chi_k)\Big] \cos\big(m\omega_0 t\big)\Big)\, .\label{eq_71}
			\end{aligned}
		\end{equation}
		This expression supplies information on the average value of  $q(t)w_k(t)$ in the NESS, as well as the amplitudes of its harmonic orders in the frequency domain. The absolute value of the constant term, |$\frac{\overline{\xi_k} q_1}{\chi_k}\, \mathcal{J}_1(\chi_k)$|, represents the average value of the $q(t) w_k(t)$ or the amplitude of $0$-th order harmonic. The absolute value of the coefficient of $\cos(m\, \omega_0 t)$ term represents twice the amplitude of $m$-th harmonic order.
		
		Next, we arrange $\omega_0 = 2\varepsilon_k -\delta$ to ensure that the phonon frequency becomes detuned from two-triplon band, namely $\omega_0\neq 2\varepsilon_k$, and conduct a similar procedure to find analytical solutions to the detuned scenario. Thus, we obtain
		\begin{equation}\label{eq_76}	
			z_k(t) =  i\,\zeta_k \, e^{i \eta_k(t)
				-\gamma_{\rm{s}} t} 		\int_{0}^{t}   \big(n_k(t') + {\textstyle \frac{3}{2}}\big)\, e^{(i\delta-\gamma_{\rm{s}}) t'} \,dt'\, .
		\end{equation} 	 and, following a course of actions, the triplon occupation of the detuned case turns out as
		\begin{equation}\label{eq_72}
			\begin{aligned}
				n_k(t) = \frac{3\, \zeta_k ^2}{2 (\gamma_{\rm s}^2 -\zeta_k^2 + \delta^2)} \big[&1+\frac{1}{2}(\frac{\gamma_{\rm s}}{\tilde{\zeta}_k}-1)e^{-(\gamma_{\rm s}+\tilde{\zeta}_k) \,t}-\frac{1}{2}(\frac{\gamma_{\rm s}}{\tilde{\zeta}_k}+1)e^{-(\gamma_{\rm s}-\tilde{\zeta}_k) \,t} \big]  \, ,
			\end{aligned}
		\end{equation}
		with the conditions that
		\begin{subequations}\label{eq_73}
			\begin{align}
				&\text{if}\,\, |\zeta_k|>|\delta|,\quad\, \text{then}\,\, \tilde{\zeta}_k=\sqrt{\zeta_k^2-\delta^2}\,,\label{eq_73a}\\
				&\text{if}\,\, |\zeta_k|<|\delta|,\quad\, \text{then}\,\, \tilde{\zeta}_k=i\sqrt{\delta^2-\zeta_k^2}\,. \label{eq_73b}
			\end{align}
		\end{subequations}
		For $|\zeta_k|>|\delta|$ case, $\gamma_{\rm{s}}>\tilde{\zeta}_k$ should also hold true to avoid negative triplon occupation and divergence of the exponential terms at $t \to \infty$ limit, whereas for $|\zeta_k|<|\delta|$ case no such condition is needed. Accordingly Eq.~\eqref{eq_72} in the NESS yields $\lim_{t \to \infty} n_k(t) = \frac{3\, \zeta_k ^2}{2 (\gamma_{\rm s}^2 -\zeta_k^2 + \delta^2)}$ which is the average triplon occupation of the detuned case. By inserting the full $n_k(t)$ expression of Eq.~\eqref{eq_72} in the Eq.~\eqref{eq_76}, $z_k(t)$ in the NESS becomes 
		\begin{equation}\label{eq_74}
			z_k(t)=\underbrace{\overline{\tilde{\xi}_k}\big[{}-\frac{\delta }{\gamma_{\rm s}}\cos\big(\eta_k(t)-\delta t\big) - \sin\big(\eta_k(t)-\delta t\big)\big]}_{v_k(t)} + i\underbrace{\overline{\tilde{\xi}_k}\big[{} \cos\big(\eta_k(t)-\delta t\big) -\frac{\delta }{\gamma_{\rm s}} \sin\big(\eta_k(t)-\delta t\big)\big]}_{w_k(t)}\,,
		\end{equation}where $\overline{\tilde{\xi}_k} = \frac{3\,\zeta_k \gamma_{\rm s}}{ 2\,(\gamma_{\rm s}^2 -\zeta_k^2 + \delta^2)}$.
		
		To find the analytical expressions of the amplitudes of each harmonic order in the detuned case, we implement the same strategy as the in-band case. Knowing that in the detuned case $\eta_k(t)=\omega_0 t+\delta t+\chi_k\sin(\omega_0 t)$, the  $\cos\big(\eta_k(t)-\delta t\big)$ and $\sin\big(\eta_k(t)-\delta t\big)$ terms are decomposed as
		\begin{subequations}
			\begin{align}
				\cos\big(\eta_k(t)-\delta t\big)=\, &\cos(\omega_0 t)\,\cos\big(\chi_k \sin(\omega_0 t)\big) -\sin(\omega_0 t)\,\sin\big(\chi_k \sin(\omega_0 t)\big)\,, \\
				\sin\big(\eta_k(t)-\delta t\big)=\, &\sin(\omega_0 t)\,\cos\big(\chi_k \sin(\omega_0 t)\big) +\cos(\omega_0 t)\,\sin\big(\chi_k \sin(\omega_0 t)\big)\,.
			\end{align}
		\end{subequations}
		By utilizing the Eqs.\eqref{eq_43} and~\eqref{eq_44}, we achieve 
		\begin{equation}
			\begin{aligned}
				&q(t)w_k(t)={} \frac{\overline{\tilde{\xi}_k} q_1 }{\chi_k} \mathcal{J}_1(\chi_k)+\frac{\overline{\tilde{\xi}_k} q_1}{\chi_k} \sum^\infty_{m=1} \widetilde{\mathcal{C}}_{k,m}\cos(m \,\omega_0 t+\tilde{\phi}_{k,m}) ,
			\end{aligned}
		\end{equation}with 
		\begin{subequations}
			\begin{align}
				&\widetilde{\mathcal{C}}_{k,m}= [(m-1)\mathcal{J}_{m-1}(\chi_k)+(-1)^{m}(m+1)\mathcal{J}_{m+1}(\chi_k)]\sqrt{1+\Big[\frac{\delta[ (m-1)\mathcal{J}_{m-1}(\chi_k)-(-1)^{m}(m+1)\mathcal{J}_{m+1}(\chi_k)]}{\gamma_{\rm{s}}[(m-1)\mathcal{J}_{m-1}(\chi_k)+(-1)^{m}(m+1)\mathcal{J}_{m+1}(\chi_k)]}\Big]^2}\,,\\
				&\tilde{\phi}_{k,m}=\arctan \Big[ \frac{\delta[(m-1)\mathcal{J}_{m-1}(\chi_k)-(-1)^{m}(m+1)\mathcal{J}_{m+1}(\chi_k)]}{\gamma_{\rm{s}}[ (m-1)\mathcal{J}_{m-1}(\chi_k)+(-1)^{m}(m+1)\mathcal{J}_{m+1}(\chi_k)]}\Big]\,.
			\end{align}
		\end{subequations}
		The half of the absolute value of $\frac{\overline{\tilde{\xi}_k} q_1}{\chi_k}\widetilde{\mathcal{C}}_{k,m}$ term corresponds to the amplitude of $m$-th harmonic order in the frequency domain of $q(t)w_k(t)$. 
		
	}
	
	\twocolumngrid

	\bibliography{bibliography}

\end{document}